\documentclass{cernyrep}
\providecommand{\OfOrd}[1]{\ensuremath{\mathcal{O}(#1)}}
\providecommand{\Lumi}[1]
    {\ensuremath{\mathscr{L}\,#1\,\text{cm}^{-2}\,\text{s}^{-1}}}
\providecommand{\EPEM}{\Pep\Pem}
\providecommand{\pT}{\ensuremath{p\sb{\scriptstyle\mathrm{T}}}}
\providecommand{\ET}{\ensuremath{E\sb{\scriptstyle\mathrm{T}}}}
\let\textsim\texttildelow
\begin{document}
\title{Trigger and data acquisition}
\author{N. Ellis}
\institute{CERN, Geneva, Switzerland}
\maketitle
\begin{abstract}
The lectures address some of the issues of triggering and data acquisition in large high-energy physics experiments. Emphasis is placed on hadron-collider experiments that present a particularly challenging environment for event selection and data collection. However, the lectures also explain how T/DAQ systems have evolved over the years to meet new challenges. Some examples are given from early experience with LHC T/DAQ systems during the 2008 single-beam operations.
\end{abstract}
\section{Introduction}

These lectures concentrate on experiments at high-energy particle
colliders, especially the general-purpose experiments at the Large
Hadron Collider (LHC) \cite{bib_LHCmachine}. These experiments represent a very challenging
case that illustrates well the problems that have to be addressed in
state-of-the-art high-energy physics (HEP) trigger and
data-acquisition (T/DAQ) systems. This is also the area in which the
author is working (on the trigger for the ATLAS experiment at LHC) and
so is the example that he knows best. However, the lectures start with
a more general discussion, building up to some examples from LEP  \cite{bib_LEPmachine} that
had complementary challenges to those of the LHC. The LEP examples
are a good reference point to
see how HEP T/DAQ systems have evolved in the last few years.

Students at this school come from various backgrounds\,---\,phenomenology,
experimental data analysis in running experiments,
and preparing for future experiments (including working
on T/DAQ systems in some cases). These lectures try to strike a
balance between making the presentation accessible to all, and going
into some details for those already familiar with T/DAQ systems.

\subsection{Definition and scope of trigger and data acquisition}

T/DAQ is the online system that selects particle interactions of
potential interest for physics analysis (trigger), and that takes
care of collecting the corresponding data from the detectors, putting
them into a suitable format and recording them on permanent storage
(DAQ). Special modes of operation need to be considered, \eg the need
to calibrate different detectors in parallel outside of normal
data-taking periods. T/DAQ is often taken to include associated tasks,
\eg run control, monitoring, clock distribution and book-keeping, all
of which are essential for efficient collection and
subsequent offline analysis of the data.

\subsection{Basic trigger requirements}

As introduced above, the trigger is responsible for selecting
interactions that are of potential interest for physics
analysis. These interactions should be selected with high efficiency,
the efficiency should be precisely known (since it enters in the
calculation of cross-sections), and there should not be biases that
affect the physics results. At the same time, a large reduction of
rate from unwanted high-rate processes may be needed to match the
capabilities of the DAQ system and the offline computing
system. High-rate processes that need to be rejected may be
instrumental backgrounds or high-rate physics processes that are not
relevant for the analyses that one wants to make. The trigger system
must also be affordable, which implies limited computing power. As a
consequence, algorithms that need to be executed at high rate must be
fast. Note that it is not always easy to achieve the above
requirements (high efficiency for signal, strong background rejection
and fast algorithms) simultaneously.

Trigger systems typically select events\footnote{The term `event' will
be discussed in Section~\ref{sec:colliderexperiments}\,---\,for now, it
may be taken to mean the record of an interaction.} according to a
`trigger menu', \ie a list of selection criteria\,---\,an event is
selected if one or more of the criteria are met. Different criteria
may correspond to different signatures for the same physics
process\,---\,redundant selections lead to high selection efficiency
and allow the efficiency of the trigger to be measured from the
data. Different criteria may also reflect the wish to concurrently
select events for a wide range of physics studies --- HEP `experiments'
(especially those with large general-purpose `detectors' or, more
precisely, detector systems) are really experimental facilities. Note
that the menu has to cover the physics channels to be studied, plus
additional data samples required to complete the analysis (\eg measure
backgrounds, and check the detector calibration and alignment).

\subsection{Basic data-acquisition requirements}

The DAQ system is responsible for the collection of data from detector
digitization systems, storing the data pending the trigger decision,
and recording data from the selected events in a suitable format. In
doing so it must avoid corruption or loss of data, and it must
introduce as little dead-time as possible (`dead-time' refers to
periods when interesting interactions cannot be selected\,---\,see
below). The DAQ system must, of course, also be affordable which, for
example, places limitations on the amount of data that can be read out
from the detectors.

\section{Design of a trigger and data-acquisition system}
\label{sec:design}

In the following a very simple example is used to illustrate some of
the main issues for designing a T/DAQ system. An attempt is made to
omit all the detail and concentrate only on the
essentials\,---\,examples from real experiments will be discussed
later.

Before proceeding to the issue of T/DAQ system design, the concept of
dead-time, which will be an important element in what follows, is
introduced. `Dead-time' is generally defined as the fraction or
percentage of total time where valid interactions could not be
recorded for various reasons. For example, there is typically a
minimum period between triggers\,---\,after each trigger the
experiment is dead for a short time.

Dead-time can arise from a number of sources, with a typical total of
up to \OfOrd{10\%}. Sources include readout and trigger dead-time, which are
addressed in detail below, operational dead-time (\eg~time to
start/stop data-taking runs), T/DAQ downtime (\eg following a
computer failure), and detector downtime (\eg following a
high-voltage trip). Given the huge investment in the accelerators and
the detectors for a modern HEP experiment, it is clearly very
important to keep dead-time to a minimum.

In the following, the design issues for a T/DAQ system are illustrated
using a very simple example. Consider an experiment that makes a
time-of-flight measurement using a scintillation-counter telescope,
read out with time-to-digital converters (TDCs), as shown in
Fig. 1. Each plane of the telescope is viewed by a photomultiplier
tube (PMT) and the resulting electronic signal is passed to a 
`discriminator' circuit that gives a digital pulse with a sharp
leading edge when a charged particle passes through the detector. The
leading edge of the pulse appears a fixed time after the particle
traverses the counter. (The PMTs and discriminators are not shown in
the figure.)

Two of the telescope planes are mounted close together, while the
third is located a considerable distance downstream giving a
measurable flight time that can be used to determine the particle's
velocity. The trigger is formed by requiring a coincidence (logical
AND) of the signals from the first two planes, avoiding triggers due
to random noise in the photomultipliers\,---\,it is very unlikely for
there to be noise pulses simultaneously from both PMTs. The time of
arrival of the particle at the three telescope planes is measured,
relative to the trigger signal, using three channels of a TDC. The
pulses going to the TDC from each of the three planes have to be
delayed so that the trigger signal, used to start the TDC (analogous
to starting a stop-watch), gets there first.

The trigger signal is also sent to the DAQ computer, telling it to
initiate the readout. Not shown in Fig.1 is logic that prevents
further triggers until the data from the TDC have been read out into
the computer\,---\,the so-called dead-time logic.

\subsection{Traditional approach to trigger and data acquisition}

The following discussion starts by presenting a `traditional' approach
to T/DAQ (as might be implemented using, for example, NIM and CAMAC
electronics modules\footnote{NIM  \cite{bib_NIM} and CAMAC \cite{bib_CAMAC} modules are electronic modules that conform to agreed standards\,---\,modules for many
functions needed in a T/DAQ system are available commercially.}, plus
a DAQ computer). Note that this approach is still widely used in small
test set-ups. The limitations of this model are described and ways of
improving on it are presented. Of course, a big HEP experiment has an
enormous number of sensor channels [up to \OfOrd{10^8} at LHC], compared to
just three in the example. However, the principles are the same, as
will be shown later.

\begin{figure}[ht]
\centering\includegraphics[width=.8\linewidth]{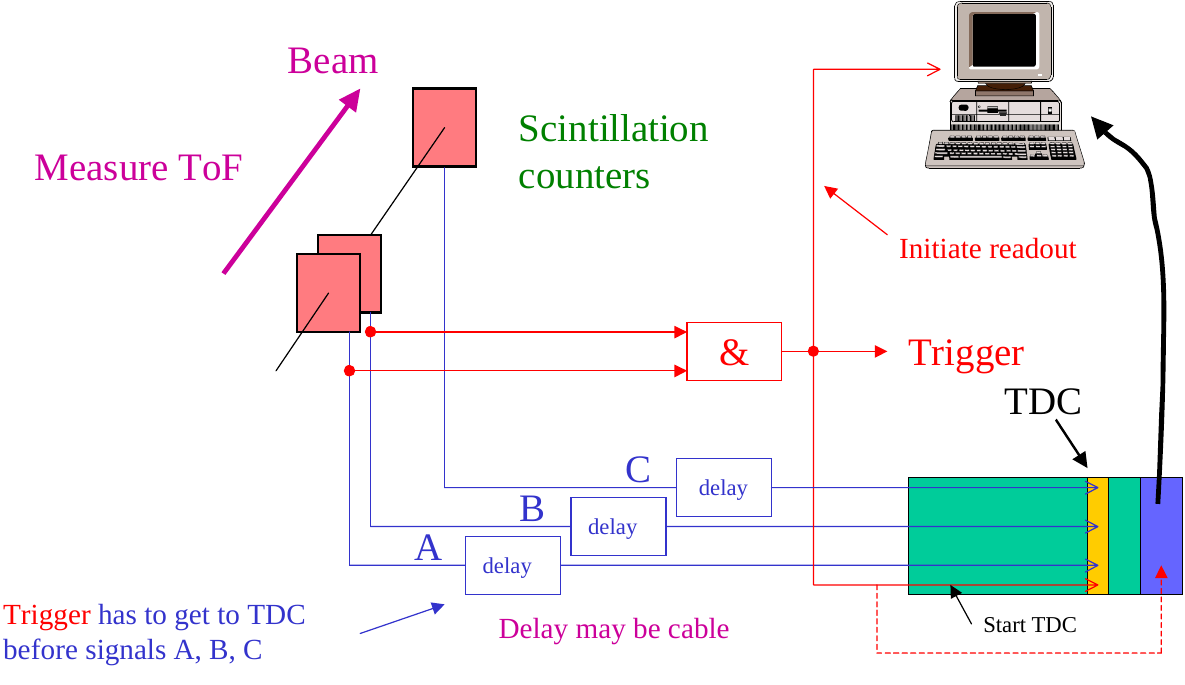}
\caption[]{Example of a simple experiment with its T/DAQ system}
\label{fig:f1}
\end{figure}

Limitations of the T/DAQ system shown in \Figure~\ref{fig:f1} are as
follows:

\begin{enumerate}
\item
  The trigger decision has to be made very quickly because the TDCs
  require a `start' signal that arrives before the signals that are
  to be digitized (a TDC module is essentially a multichannel digital
  stop-watch). The situation is similar with traditional
  analog-to-digital converters (ADCs) that digitize the magnitude of
  a signal arriving during a `gate' period, \eg the electric charge
  in an analog pulse\,---\,the gate has to start before the pulse
  arrives.
\item
 The readout of the TDCs by the computer may be quite slow, which
  implies a significant dead-time if the trigger rate is high. This
  limitation becomes much more important in larger systems, where many
  channels have to be read out for each event. For example, if 1000
  channels have to be read out with a readout time of 1\Uus{} per
  channel (as in CAMAC), the readout time per event is 1~ms which
  excludes event rates above 1\UkHz.
\end{enumerate}

\begin{figure}[ht]
\centering\includegraphics[width=.8\linewidth]{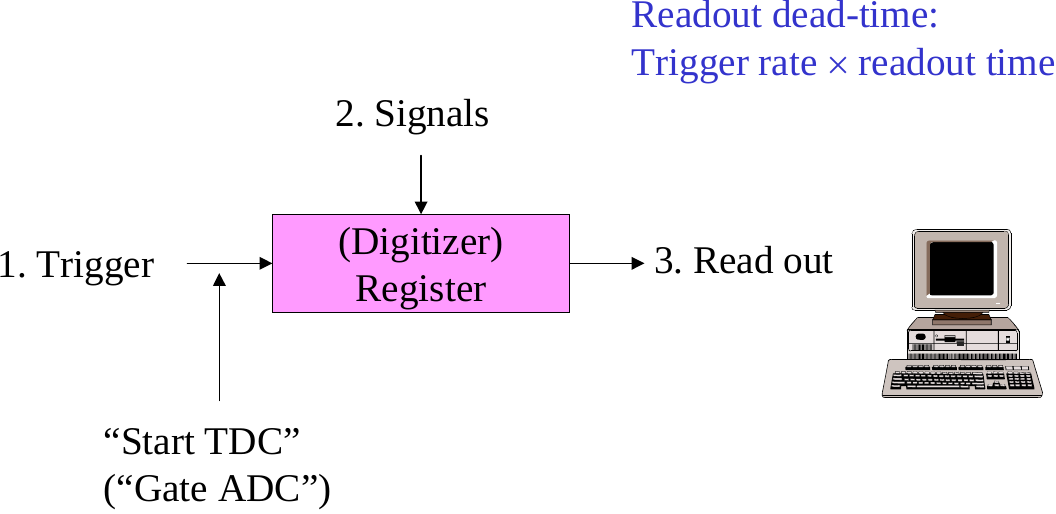}
\caption[]{Readout model in the `traditional' approach}
\label{fig:f2}
\end{figure}

The `readout model' of this traditional approach to T/DAQ is
illustrated in \Fref{fig:f2}, which shows the sequence of
actions\,---\,arrival of the trigger, arrival of the detector signals
(followed by digitization and storage in a data register in the TDC),
and readout into the DAQ computer. Since no new trigger can be
accepted until the readout is complete, the readout dead-time is given
by the product of the trigger rate and the readout time.

\newpage
\subsection{Local buffer}

\begin{figure}[ht]
\centering\includegraphics[width=.8\linewidth]{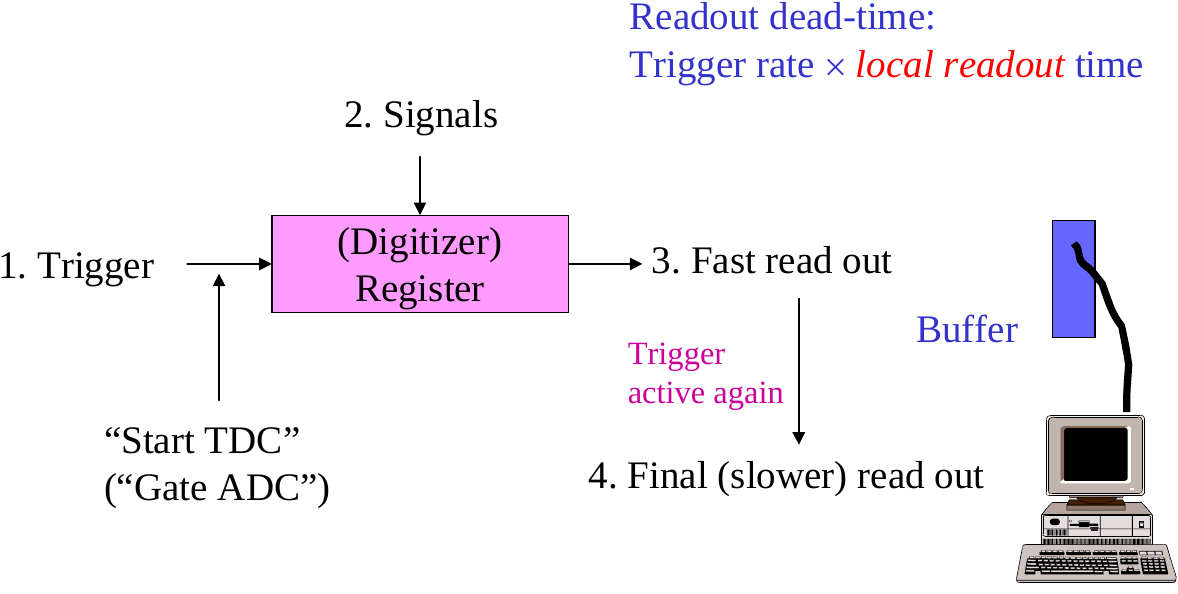}
\caption[]{Readout system with local buffer memory}
\label{fig:f3}
\end{figure}

The traditional approach described above can be improved by adding a
local `buffer' memory into which the data are moved rapidly
following a trigger, as illustrated in \Figure~\ref{fig:f3}. This fast readout
reduces the dead-~time, which is now given by the product of the
trigger rate and the local readout time. This approach is particularly
useful in large systems, where the transfer of data can proceed in
parallel with many local buffers (\eg one local buffer for each crate of
electronics)\,---\,local readout can remain fast even in a large
system. Also, the data may be moved more quickly into the local buffer
within the crate than into the DAQ computer. Note that the dashed line
in the bottom, right-hand part of Fig. 1 indicates this extension to
the traditional approach\,---\,the trigger signal is used to initiate
the local readout within the crate.

The addition of a local buffer reduces the effective readout time, but
the requirement of a fast trigger still remains. Signals have to be
delayed until the trigger decision is available at the
digitizers. This is not easy to achieve, even with very simple trigger
logic\,---\,typically one relies on using fast (air-core) cables for
trigger signals with the shortest possible routing so that the trigger
signals arrive before the rest of the signals (which follow a longer
routing on slower cables). It is not possible to apply complex
selection criteria on this time-scale.

\subsection{Multi-level triggers}

It is not always possible to simultaneously meet the physics
requirements (high efficiency, high background rejection) and achieve
an extremely short trigger `latency' (time to form the trigger
decision and distribute it to the digitizers). A solution is to
introduce the concept of multi-level triggers, where the first level
has a short latency and maintains high efficiency, but only has a
modest rejection power. Further background rejection comes from the
higher trigger levels that can be slower. Sometimes the very fast
first stage of the trigger is called the `pre-trigger'\,---\,it may
be sufficient to signal the presence of minimal activity in the
detectors at this stage.

\begin{figure}[ht]
\centering\includegraphics[width=.8\linewidth]{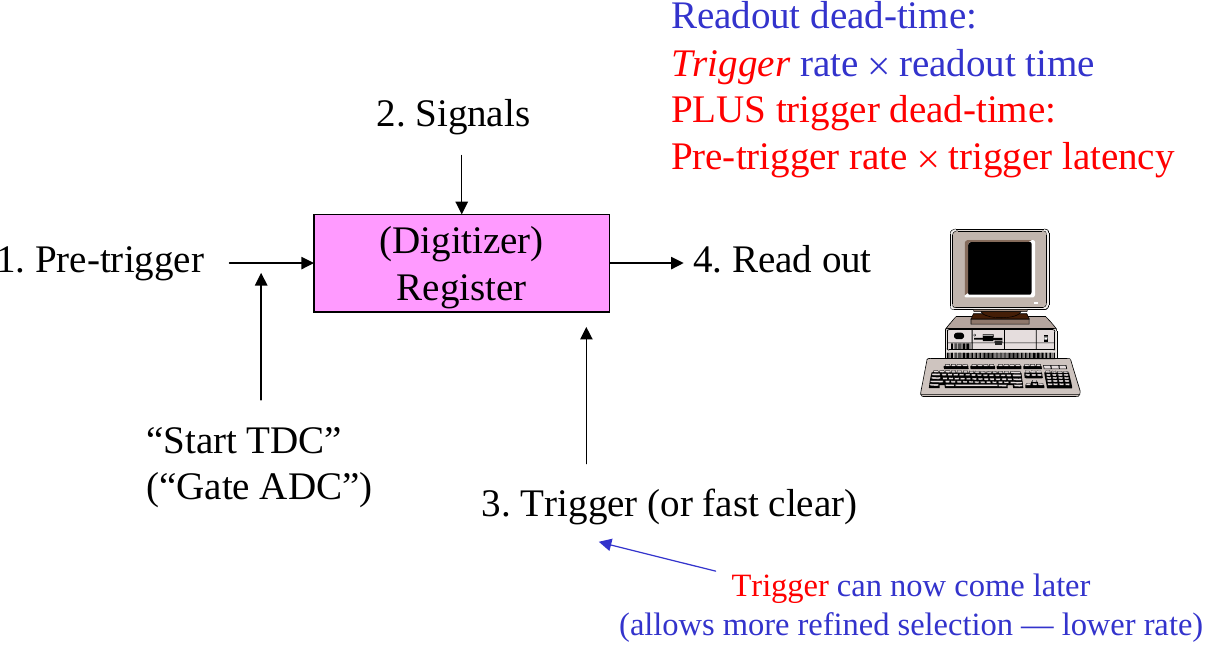}
\caption[]{Readout system with pre-trigger and fast clear}
\label{fig:f4}
\end{figure}

The use of a pre-trigger is illustrated in \Fref{fig:f4}. Here the
pre-trigger is used to provide the start signal to the TDCs (and to
gate ADCs, \etc), while the main trigger (which can come later and can
therefore be based on more complex calculations) is used to initiate
the readout. In cases where the pre-trigger is not confirmed by the
main trigger, a `fast clear' is used to re-activate the digitizers
(TDCs, ADCs, \etc).

Using a pre-trigger (but without using a local buffer for now), the
dead-time has two components. Following each pre-trigger there is a
dead period until the trigger or fast clear is issued (defined here as
the trigger latency). For the subset of pre-triggers that give rise to
a trigger, there is an additional dead period given by the readout
time. Hence, the total dead-time is given by the product of the
pre-trigger rate and the trigger latency, added to the product of the
trigger rate and the readout time.

\begin{figure}[ht]
\centering\includegraphics[width=.8\linewidth]{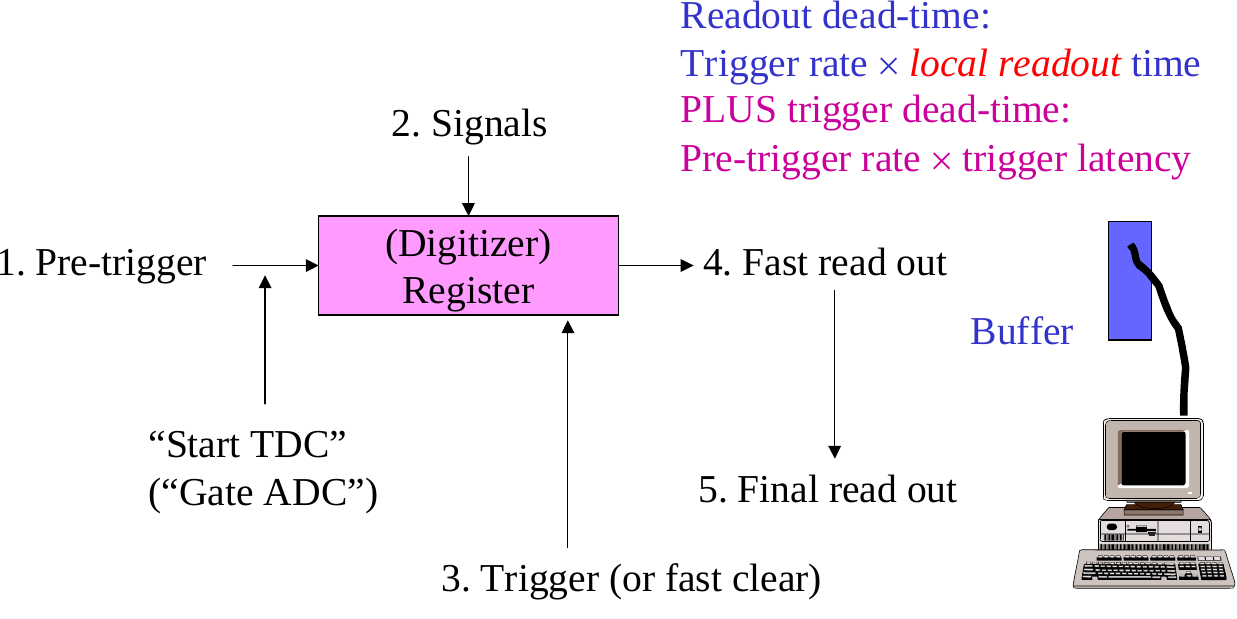}
\caption[]{Readout system using both pre-trigger and local buffer}
\label{fig:f5}
\end{figure}

The two ingredients\,---\,use of a local buffer and use of a
pre-trigger with fast clear\,---\,can be combined as shown in
\Figure~\ref{fig:f5}, further reducing the dead-time. Here the total
dead-time is given by the product of the pre-trigger rate and the
trigger latency, added to the product of the trigger rate and the
local readout time.

\subsection{Further improvements}

The idea of multi-level triggers can be extended beyond having two
levels (pre-trigger and main trigger). One can have a series of
trigger levels that progressively reduce the rate. The efficiency for
the desired physics must be kept high at all levels since rejected
events are lost forever. The initial levels can have modest rejection
power, but they must be fast since they see a high input rate. The
final levels must have strong rejection power, but they can be slower
because they see a much lower rate (thanks to the rejection from the
earlier levels).
\newpage
In a multi-level trigger system, the total dead-time can be written as
the sum of two parts: the trigger dead-time summed over trigger
levels, and the readout dead-time. For a system with \emph{N} levels, this
can be written
\begin{displaymath}
( \sum^N_{i=2} R_{i-1} \times L_i ) + R_N \times T _{\mathrm{LRO}} 
\end{displaymath}
where $R_i$ is the rate after the $i^{\mathrm{th}}$ trigger level,
$L_i$ is the latency of the $i^{\mathrm{th}}$ trigger level, and $T
_{\mathrm{LRO}}$ is the local readout time. Note that $R_1$
corresponds to the pre-trigger rate.

In the above, two implicit assumptions have been made: (1) that all
trigger levels are completed before the readout starts, and (2) that
the pre-trigger (\ie the lowest-level trigger) is available by the
time the first signals from the detector arrive at the digitizers. The
first assumption results in a long dead period for some
events\,---\,those that survive the first (fast) levels of
selection. The dead-time can be reduced by moving the data into
intermediate storage after the initial stages of trigger selection,
after which further low-level triggers can be accepted (in parallel
with the execution of the later stages of trigger selection on the
first event). The second assumption can also be avoided, \eg in
collider experiments with bunched beams as discussed below.

In the next section, aspects of particle colliders that affect the
design of T/DAQ systems are introduced. Afterwards, the discussion
returns to readout models and dead-time, considering the example of LEP
experiments.

\section{Collider experiments}
\label{sec:colliderexperiments}

In high-energy particle colliders (HERA, LEP, LHC, Tevatron), the
particles in the counter-rotating beams are bunched. Bunches of
particles cross at regular intervals and interactions occur only
during the bunch crossings. Here the trigger has the job of selecting
the \emph{bunch crossings} of interest for physics analysis, \ie those
containing interactions of interest.

In the following notes, the term `event' is used to refer to the
record of all the products from a given bunch crossing (plus any
activity from other bunch crossings that gets recorded along with
this). Be aware (and beware!)\,---\,the term `event' is not uniquely
defined! Some people use the term `event' for the products of a
single interaction between incident particles. Note that many people
use `event' interchangeably to mean different things.

In {\EPEM} colliders, the interaction rate is very small compared to
the bunch-crossing rate (because of the low {\EPEM}
cross-section). Generally, selected events contain just one
interaction\,---\,\ie the event is generally a single
interaction. This was the case at LEP and also at the
{\Pe--\Pp} collider HERA. In contrast, at LHC with the design
luminosity \Lumi{\text{~of~}10^{34}}  for proton beams,
each bunch crossing will contain
on average about 25 interactions as discussed below. This means that
an interaction of interest, \eg one that produced $\PH \rightarrow
\PZ\PZ \rightarrow \Pep\Pem\Pep\Pem$, will be recorded together with
~25 other proton--proton interactions that occurred in the same bunch
crossing. The interactions that make up the `underlying event' are
often called `minimum-bias' interactions because they are the ones
that would be selected by a trigger that selects interactions in an
unbiased way. The presence of additional interactions that are
recorded together with the one of interest is sometimes referred to as
`pile-up'.

A further complication is that particle detectors do not have an
infinitely fast response time\,---\,this is analogous to the exposure
time of a camera. If the `exposure time' is shorter than the
bunch-crossing period, the event will contain only information from
the selected bunch crossing. Otherwise, the event will contain,
in addition, any activity from neighbouring bunches. In
{\EPEM} colliders (\eg LEP) it is very unlikely for there to be any
activity in nearby bunch crossings, which allows the use of slow
detectors such as the time projection chamber (TPC). This is also true
at HERA and in the ALICE experiment~\cite{bib_ALICE} at LHC that will study
heavy-ion collisions at much lower luminosities than in the
proton--proton case.

The bunch-crossing period for proton--proton collisions at LHC will be
only 25\Uns{} (corresponding to a 40\UMHz~rate). At the design luminosity the interaction
rate will be \OfOrd{10^9}\UHz~and, even with the short bunch-crossing
period, there will be an average of about 25 interactions per bunch
crossing. Some detectors, for example the ATLAS silicon tracker,
achieve an exposure time of less than 25\Uns, but many do not. For
example, pulses from the ATLAS liquid-argon calorimeter extend over
many bunch crossings.

The instrumentation for the LHC experiments is described in the lecture notes of Jordan Nash from this School~\cite{bib_Nash}. The Particle Data Group's Review of Particle Physics~\cite{bib_PDG} includes much useful information, including summaries of the parameters of various particle colliders.

\section{Design of a trigger and data-acquisition system for LEP}
\label{sec:designlep}

Let us now return to the discussion of designing a T/DAQ system,
considering the case of experiments at LEP (ALEPH~\cite{bib_ALEPH},
DELPHI~\cite{bib_DELPHI}, L3~\cite{bib_L3}, and OPAL~\cite{bib_OPAL}), and building on
the model developed in Section~\ref{sec:design}.

\begin{figure}[t]
\centering\includegraphics[width=.6\linewidth]{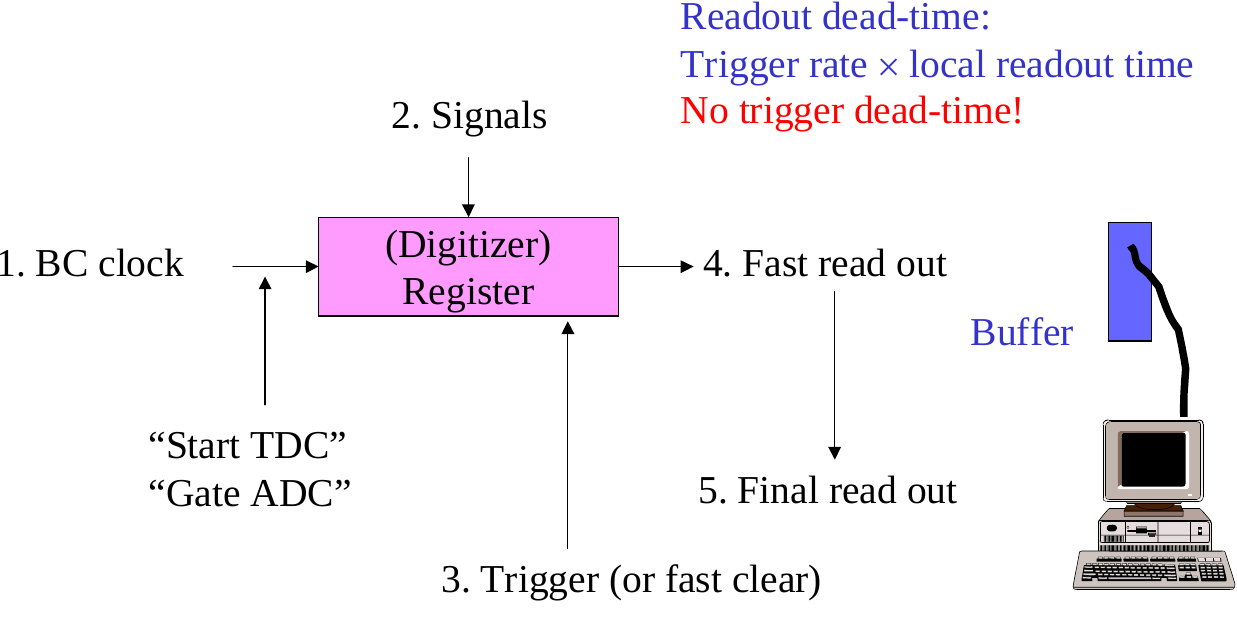}
\caption[]{Readout system using bunch-crossing (BC) clock and fast clear}
\label{fig:f6}
\end{figure}

\subsection{Using the bunch-crossing signal as a `pre-trigger'}

If the time between bunch crossings (BCs) is reasonably long, one can
use the clock that signals when bunches of particles cross as the
pre-trigger. The first-level trigger can then use the time between
bunch crossings to make a decision, as shown in
\Figure~\ref{fig:f6}. For most crossings the trigger will reject the
event by issuing a fast clear\,---\,in such cases no dead-time is
introduced. Following an `accept' signal, dead-time will be introduced
until the data have been read out (or until the event has been
rejected by a higher-level trigger). This is the basis of the model
that was used at LEP, where the bunch-crossing interval of
22~{\textmu}s (11~{\textmu}s in eight-bunch mode) allowed
comparatively complicated trigger processing (latency of a few
microseconds). Note that there is no first-level trigger dead-time
because the decision is made during the interval between bunch
crossings where no interactions occur. As discussed below, the trigger
rates were reasonably low (very much less than the BC rate), giving
acceptable dead-time due to the second-level trigger latency and the
readout.

\begin{figure}[t]
\centering\includegraphics[width=.8\linewidth]{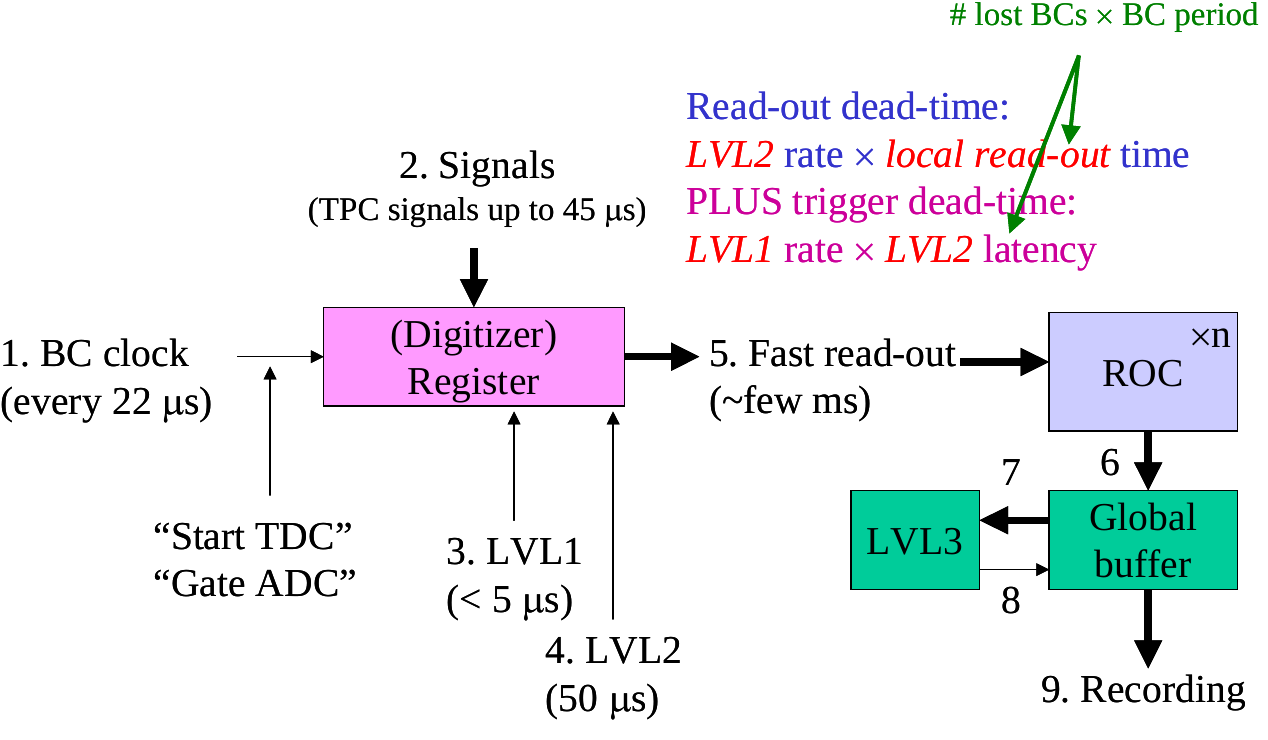}
\caption[]{LEP readout model (ALEPH)}
\label{fig:f7}
\end{figure}

In the following, the readout model used at LEP is illustrated by
concentrating on the example of ALEPH~\cite{bib_ALEPH}\footnote{The author
was not involved in any of the LEP experiments. In these lectures the
example of ALEPH is used to illustrate how triggers and
data acquisition were implemented at LEP; some numbers from DELPHI are
also presented. The T/DAQ systems in all of the LEP experiments were
conceptually similar.}. \Figure[b]~\ref{fig:f7} shows the readout
model, using the same kind of block diagram as presented in Section
2. The BC clock is used to start the TDCs and generate the gate for
the ADCs, and a first-level (LVL1) trigger decision arrives in less
than 5 {\textmu}s so that the fast clear can be completed prior to the
next bunch crossing. For events retained by LVL1, a more sophisticated
second-level (LVL2) trigger decision is made after a total of about 50
{\textmu}s. Events retained by LVL2 are read out to local buffer
memory (within the readout controllers or `ROCs'), and then passed to
a global buffer. There is a final level of selection (LVL3) before
recording the data on permanent storage for offline analysis.

For readout systems of the type shown in \Figure~\ref{fig:f7}, the
total dead-time is given by the sum of two components\,---\,the
trigger dead-time and the readout dead-time.

The trigger dead-time is evaluated by counting the number of BCs that
are lost following each LVL1 trigger, then calculating the product of
the LVL1 trigger rate, the number of lost BCs and the BC period. Note
that the effective LVL2 latency, given by the number of lost BCs and
the BC period, is less than (or equal to) the true LVL2 latency.

The readout dead-time is given by the product of the LVL2 trigger rate
and the time taken to perform local readout into the ROCs. Strictly
speaking, one should also express this dead-time in terms of the
number of BCs lost after the LVL2 trigger, but since the readout time
is much longer than the BC period the difference is unimportant. Note
that, as long as the buffers in the ROCs and the global buffers do not
fill up, no additional dead-time is introduced by the final readout
and the LVL3 trigger.

Let us now look quantitatively at the example of the DELPHI
experiment for which the T/DAQ system was similar to that described
above for ALEPH. Typical numbers for LEP-II\footnote{LEP-II refers to the period when LEP operated at high energy, after the upgrade of the RF system.} are shown in
\mbox{\Table~\ref{tab:tdabparameters}~\cite{bib_DELPHI}}.

\subsection{Data acquisition at LEP}

\begin{figure}[p]
\centering\includegraphics[width=.6\linewidth]{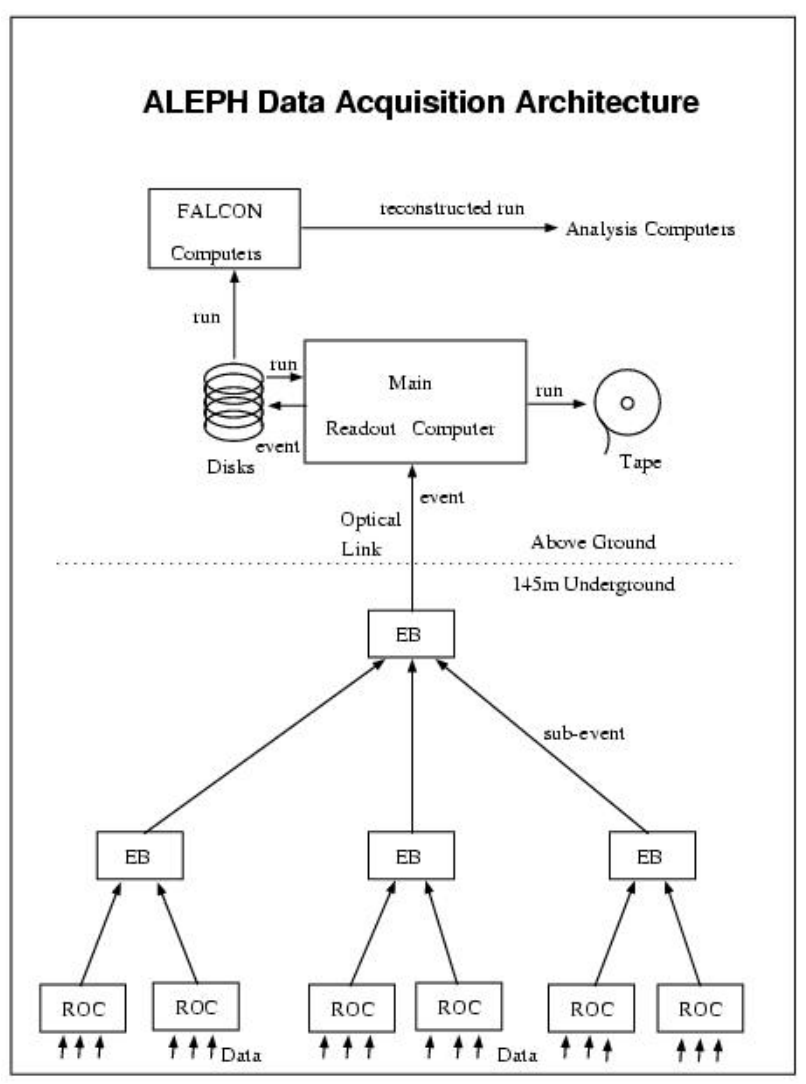}
\caption[]{ALEPH data-acquisition architecture}
\label{fig:f8}

\medskip

\centering\includegraphics[width=.8\linewidth]{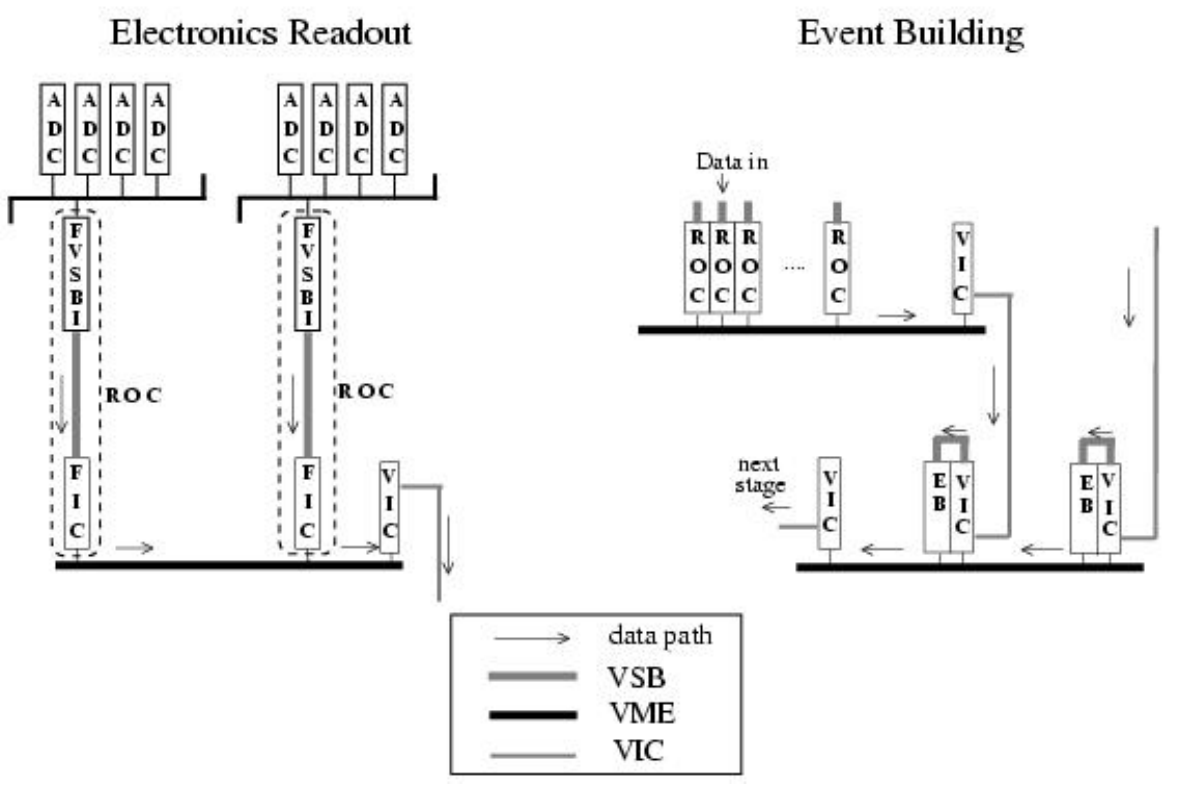}
\caption[]{Event building in ALEPH}
\label{fig:f9}
\end{figure}

Let us now continue our examination of the example of the ALEPH T/DAQ
system. Following a LVL2 trigger, events were read out locally and in
parallel within the many readout crates\,---\,once the data had been
transferred within each crate to the ROC, further LVL1 and LVL2
triggers could be accepted. Subsequently, the data from the readout
crates were collected by the main readout computer, `building' a
complete event. As shown in \Figure~\ref{fig:f8}, event building was
performed in two stages: an event contained a number of sub-events,
each of which was composed of several ROC data blocks. Once a complete
event was in the main readout computer, the LVL3 trigger made a final
selection before the data were recorded.

\begin{table}
\caption{Typical T/DAQ parameters for the DELPHI experiment at LEP-II}
\label{tab:tdabparameters}
\renewcommand{\arraystretch}{1.1}%
\centering
\begin{tabular}{@{}l@{\qquad}l@{}}                             
\hline\hline
\textbf{Quantity}& 
   \textbf{Value}                                  \\\hline
LVL1 rate        &
   \textsim\,500--1000\UHz~(instrumental background) \\
LVL2 rate        &
   6--8\UHz                                         \\
LVL3 rate        &
   4--6\UHz                                         \\
LVL2 latency     &
   38\Uus{} (1 lost BC $\Rightarrow$ 22\Uus~effective) \\
Local readout time &
   \textsim\,2.5\Ums                                 \\
Readout dead-time  &
   \textsim\,7\UHz~\texttimes~2.5 $\cdot$ 10\textsuperscript{-3}\Us{} = 1.8\%\\
Trigger dead-time  &
   \textsim\,750\UHz~\texttimes~22 $\cdot$ 10\textsuperscript{-6}\Us{} = 1.7\%\\
Total dead-time    & 
   \textsim\,3--4\%                                 \\
\hline\hline
\end{tabular}
\end{table}

The DAQ system used a hierarchy of computers\,---\,the local ROCs in
each crate; event builders (EBs) for sub-events; the main EB; the main
readout computer. The ROCs performed some data processing (\eg
applying calibration algorithms to convert ADC values to energies) in
addition to reading out the data from ADCs, TDCs, \etc (Zero
suppression was already performed at the level of the digitizers where
appropriate.) The first layer of EBs combined data read out from the
ROCs of individual sub-detectors into sub-events; then the main EB
combined the sub-events for the different sub-detectors. Finally, the
main readout computer received full events from the main EB, performed
the LVL3 trigger selection, and recorded selected events for
subsequent analysis.

As indicated in \Fref{fig:f9}, event building was bus based\,---\,each
ROC collected data over a bus from the digitizing electronics; each
sub-detector EB collected data from several ROCs over a bus; the main
EB collected data from the sub-detector EBs over a bus. As a
consequence, the main EB and the main readout computer saw the full
data rate prior to the final LVL3 selection. At LEP this was
fine\,---\,with an event rate after LVL2 of a few hertz and an event size
of~100~kbytes, the data rate was a few hundred kilobytes per second, much less than
the available bandwidth (\eg ~40~Mbytes/s maximum on VME bus~\cite{bib_VME}).

\subsection{Triggers at LEP}

The triggers at LEP aimed to select any {\EPEM} annihilation event
with a visible final state, including events with little visible
energy, plus some fraction of two-photon events, plus Bhabha
scattering events. Furthermore, they aimed to select most events by
multiple, independent signatures so as to maximize the trigger
efficiency and to allow the measurement of the efficiency from the
data. The probability for an event to pass trigger A or trigger B is
\textsim\,$1 - \delta_\text{A}\delta_\text{B}$, where $\delta_\text{A}$ and
$\delta_\text{B}$ are
the individual trigger inefficiencies, which is very close to unity
for small $\delta$. Starting from a sample of events selected with
trigger A, the efficiency of trigger B can be estimated as the
fraction of events passing trigger B in addition. Note that in the
actual calculations small corrections were applied for correlations
between the trigger efficiencies.

\section{Towards the LHC}

In some experiments it is not practical to make a trigger in the time
between bunch crossings because of the short BC period\,---\,the BC
interval is 396~ns at Tevatron-II\footnote{Tevatron-II refers to the Tevatron collider after the luminosity upgrade.}, 96~ns at HERA and 25~ns at LHC. In
such cases the concept of `pipelined' readout has to be introduced
(also pipelined LVL1 trigger processing). Furthermore, in experiments
at high-luminosity hadron colliders the data rates after the LVL1
trigger selection are very high, and new ideas have to be introduced
for the high-level triggers (HLTs) and DAQ\,---\,in particular, event
building has to be based on data networks and switches rather than
data buses.

\subsection{Pipelined readout}

\begin{figure}[h]
\centering\includegraphics[width=.4\linewidth]{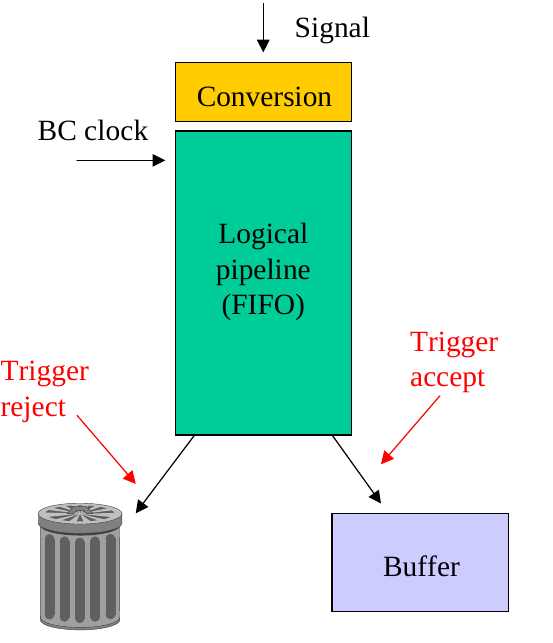}
\caption[]{Example of pipelined readout}
\label{fig:f11}
\end{figure}

In pipelined readout systems (see \Figure~\ref{fig:f11}), the
information from each BC, for each detector element, is retained
during the latency of the LVL1 trigger (several {\textmu}s). The
information may be retained in several forms\,---\,analog levels
(held on capacitors); digital values (\eg ADC results); binary values
(\ie hit or no hit). This is done using a logical `pipeline', which may
be implemented using a first-in, first-out (FIFO) memory circuit. Data
reaching the end of the pipeline are either discarded or, in the case of a
trigger accept decision, moved to a secondary buffer memory (small
fraction of BCs).

\begin{figure}[h]
\centering\includegraphics[width=.74\linewidth]{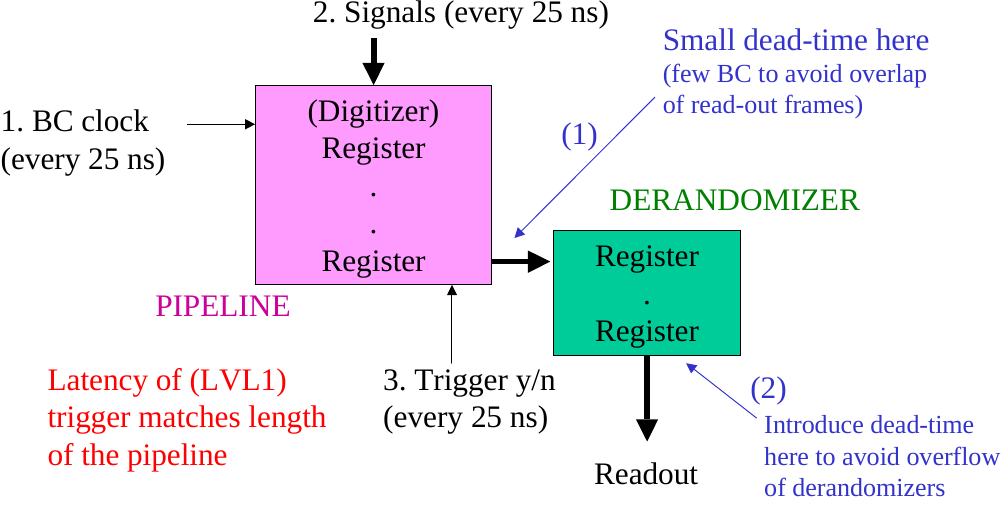}
\caption[]{Pipelined readout with derandomizer at the LHC}
\label{fig:f12}
\end{figure}

Pipelined readout systems will be used in the LHC experiments (they
have already been used in experiments at
HERA~\cite{bib_H1,bib_ZEUS} and the Tevatron~\cite{bib_CDF,bib_D0}, but the
demands at LHC are even greater because of the short BC period). A typical
LHC pipelined readout system is illustrated in \Figure~\ref{fig:f12},
where the digitizer and pipeline are driven by the 40\UMHz{} BC clock. A
LVL1 trigger decision is made for each bunch crossing (\ie every
25~ns), although the LVL1 latency is several microseconds\,---\,the
LVL1 trigger must concurrently process many events (this is achieved
by using pipelined trigger processing as discussed below).

The data for events that are selected by the LVL1 trigger are
transferred into a `derandomizer'\,---\,a memory that can accept the
high instantaneous input rate (\ie one word per 25~ns) while being
read out at the much lower average data rate (determined by the LVL1
trigger rate rather than the BC rate). In principle no dead-time needs
to be introduced in such a system. However, in practice, data are
retained for a few BCs around the one that gave rise to the trigger,
and a dead period of a few BCs is introduced to ensure that the same
data do not have to be accessed for more than one trigger. Dead-time
must also be introduced to prevent the derandomizers from overflowing,
\eg where, due to a statistical fluctuation, many LVL1 triggers arrive
in quick succession. The dead-time from the first of these sources can
be estimated as follows (numbers from ATLAS): taking a LVL1 trigger
rate of 75\UkHz{} and 4 dead BCs following each LVL1 trigger gives
$75\UkHz \times 4 \times 25\Uns = 0.75 \%$. The dead-time from the
second source depends on the size of the derandomizer and the speed
with which it can be emptied\,---\,in ATLAS the requirements are $<
1\%$ dead-time for a LVL1 rate of 75\UkHz{} ($< 6\%$ for 100\UkHz).

\begin{figure}[ht]
\centering\includegraphics[width=.6\linewidth]{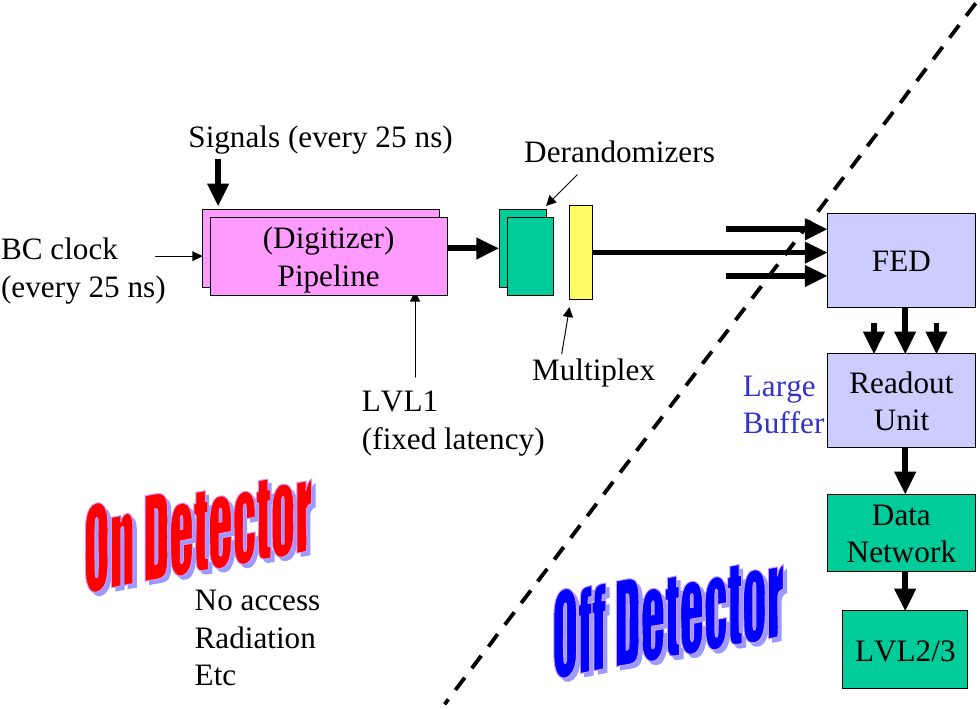}
\caption[]{Location of readout components in CMS}
\label{fig:f13}
\end{figure}

Some of the elements of the readout chain in the LHC experiments have
to be mounted on the detectors (and hence are totally inaccessible
during running of the machine and are in an environment with high
radiation levels). This is shown for the case of CMS in
\Figure~\ref{fig:f13}.

There are a variety of options for the placement of digitization in
the readout chain, and the optimum choice depends on the
characteristics of the detector in question. Digitization may be
performed on the detector at 40\UMHz{} rate, prior to a digital pipeline
(\eg CMS calorimeter). Alternatively, it may be done on the detector
after multiplexing signals from several analog pipelines (\eg ATLAS
EM calorimeter)\,---\,here the digitization rate can be lower, given
by the LVL1 trigger rate multiplied by the number of signals to be
digitized per trigger. Another alternative (\eg CMS tracker) is to
multiplex analog signals from the pipelines over analog links, and
then to perform the digitization off-detector.

\subsection{Pipelined LVL1 trigger}

As discussed above, the LVL1 trigger has to deliver a new decision
every BC, although the trigger latency is much longer than the BC
period; the LVL1 trigger must concurrently process many events. This
can be achieved by `pipelining' the processing in custom trigger
processors built using modern digital electronics. The key ingredients
in this approach are to break the processing down into a series of
steps, each of which can be performed within a single BC period, and
to perform many operations in parallel by having separate processing
logic for each calculation. Note that in such a system the latency of
the LVL1 trigger is fixed\,---\,it is determined by the number of
steps in the calculation, plus the time taken to move signals and data
to, from and between the components of the trigger system (\eg
propagation delays on cables).

\begin{figure}[ht]
\centering\includegraphics[width=.8\linewidth]{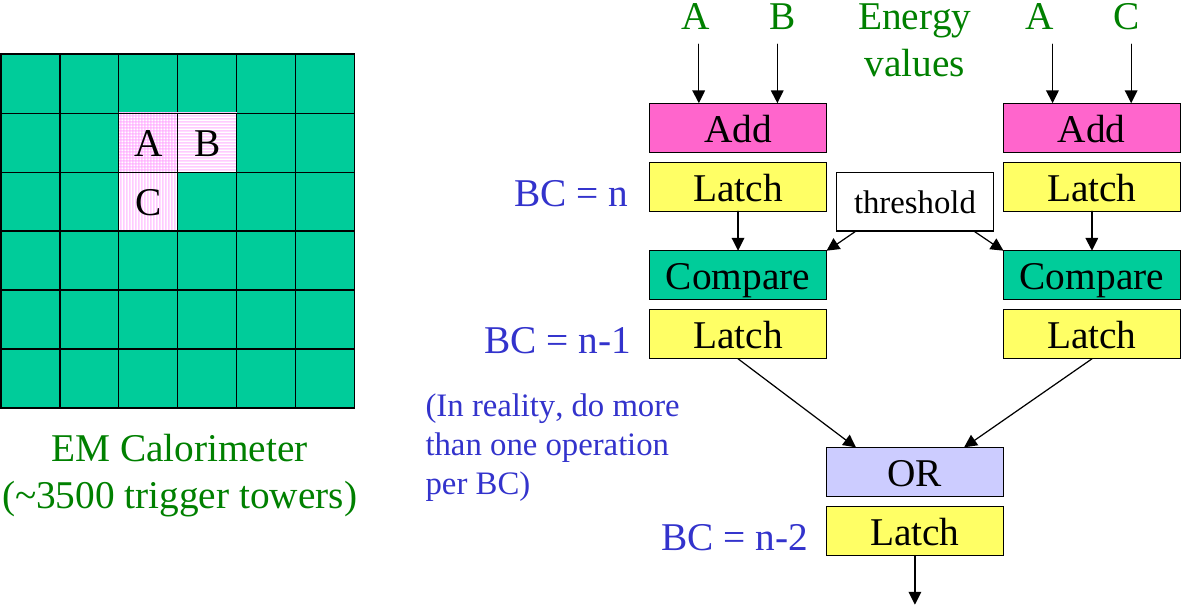}
\caption[]{Illustration of pipelined processing}
\label{fig:f14}
\end{figure}

Pipelined trigger processing is illustrated in
\Figure~\ref{fig:f14}\,---\,as will be seen later, this example
corresponds to a (very small) part of the ATLAS LVL1 calorimeter
trigger processor. The drawing on the left of \Figure~\ref{fig:f14} depicts the
EM calorimeter as a grid of `towers' in $\eta\text{--}\phi$ space
($\eta$ is pseudorapidity, $\phi$~is azimuth angle). The logic shown
on the right determines if the energy deposited in a horizontal or
vertical pair of towers in the region [A, B, C] exceeds a threshold.
In each 25~ns period, data from one layer of `latches' (memory
registers) are processed through the next step in the processing
`pipe', and the results are captured in the next layer of
latches. Note that, in the real system, such logic has to be performed
in parallel for \textsim\,3500 positions of the reference tower; the
tower `A' could be at any position in the calorimeter. In practice,
modern electronics is capable of doing more than a simple add or
compare operation in 25~ns, so there is more logic between the latches
than in this illustration.

\begin{figure}[ht]
\centering\includegraphics[width=.6\linewidth]{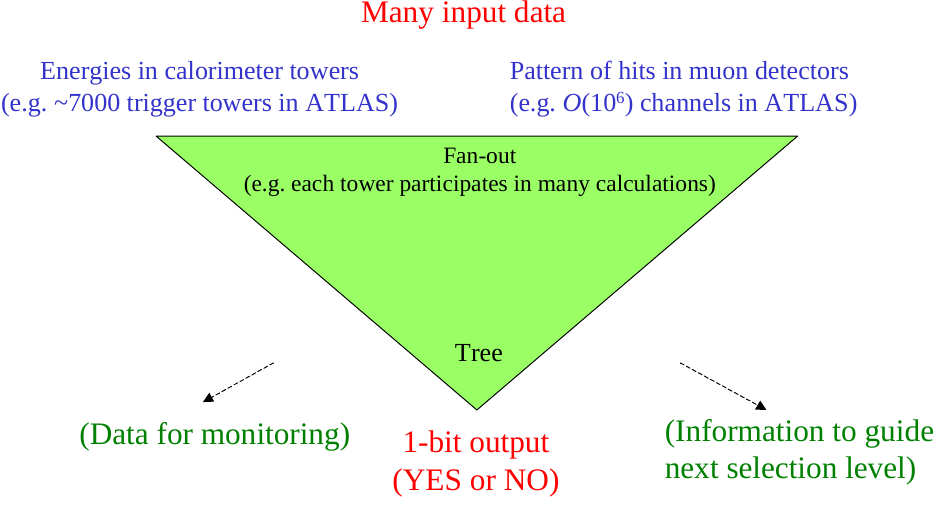}
\caption[]{LVL1 data flow}
\label{fig:f15}
\end{figure}

The amount of data to be handled varies with depth in the processing
pipeline, as indicated in \Figure~\ref{fig:f15}. Initially the amount of data
expands compared to the raw digitization level since each datum
typically participates in several operations\,---\,the input data need
to be `fanned out' to several processing elements. Subsequently the
amount of data decreases as one moves further down the processing
tree. The final trigger decision can be represented by a single bit of
information for each BC\,---\,yes or no (binary 1 or 0). Note that, in
addition to the trigger decision, the LVL1 processors produce a lot of
data for use in monitoring the system and to guide the higher levels
of selection.

Although they have not been discussed in these lectures because of time
limitations, some fixed-target experiments have very challenging T/DAQ
requirements. Some examples can be found in Refs.~\cite{bib_NA48,bib_HERA-B}.

\section{High-level triggers and data acquisition at the LHC}

In the LHC experiments, data are transferred after a LVL1 trigger
accept decision to large buffer memories\,---\,in normal operation the
subsequent stages should not introduce further dead-time. At this
point in the readout chain, the data rates are still massive. An event
size of \textsim\,1~Mbyte (after zero suppression or data compression)
at \textsim\,100\UkHz{} event rate gives a total bandwidth of
\textsim\,100~Gbytes/s (\ie~\textsim\,800~Gbits/s). This is far beyond
the capacity of the bus-based event building of LEP. Such high data
rates will be dealt with by using network-based event building and by
only moving a subset of the data.

\begin{figure}[ht]
\centering\includegraphics[width=.8\linewidth]{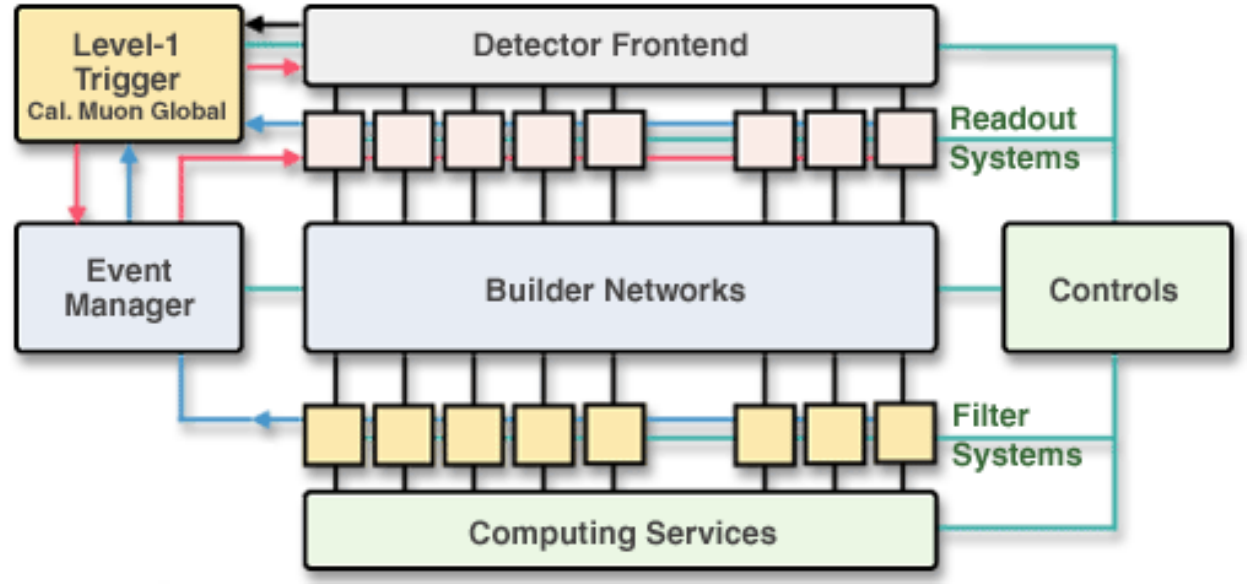}
\caption[]{CMS event builder}
\label{fig:f19}
\end{figure}

Network-based event building is illustrated in \Figure~\ref{fig:f19}
for the example of CMS. Data are stored in the readout systems until
they have been transferred to the filter systems [associated with
high-level trigger (HLT) processing], or until the event is
rejected. Note that no node in the system sees the full data
rate\,---\,each readout system covers only a part of the
detector and each filter system deals with only a fraction of the
events.

The LVL2 trigger decision can be made without accessing or processing
all of the data. Substantial rejection can be made with respect to
LVL1 without accessing the inner-tracking detectors\,---\,calorimeter
triggers can be refined using the full-precision, full-granularity
calorimeter information; muon triggers can be refined using the
high-precision readout from the muon detectors. It is therefore only
necessary to access the inner-tracking data for the subset of events
that pass this initial selection. ATLAS and CMS both use this
sequential selection strategy. Nevertheless, the massive data rates
pose problems even for network-based event building, and different
solutions have been adopted in ATLAS and CMS to address this.

\begin{figure}[ht]
\centering\includegraphics[width=.8\linewidth]{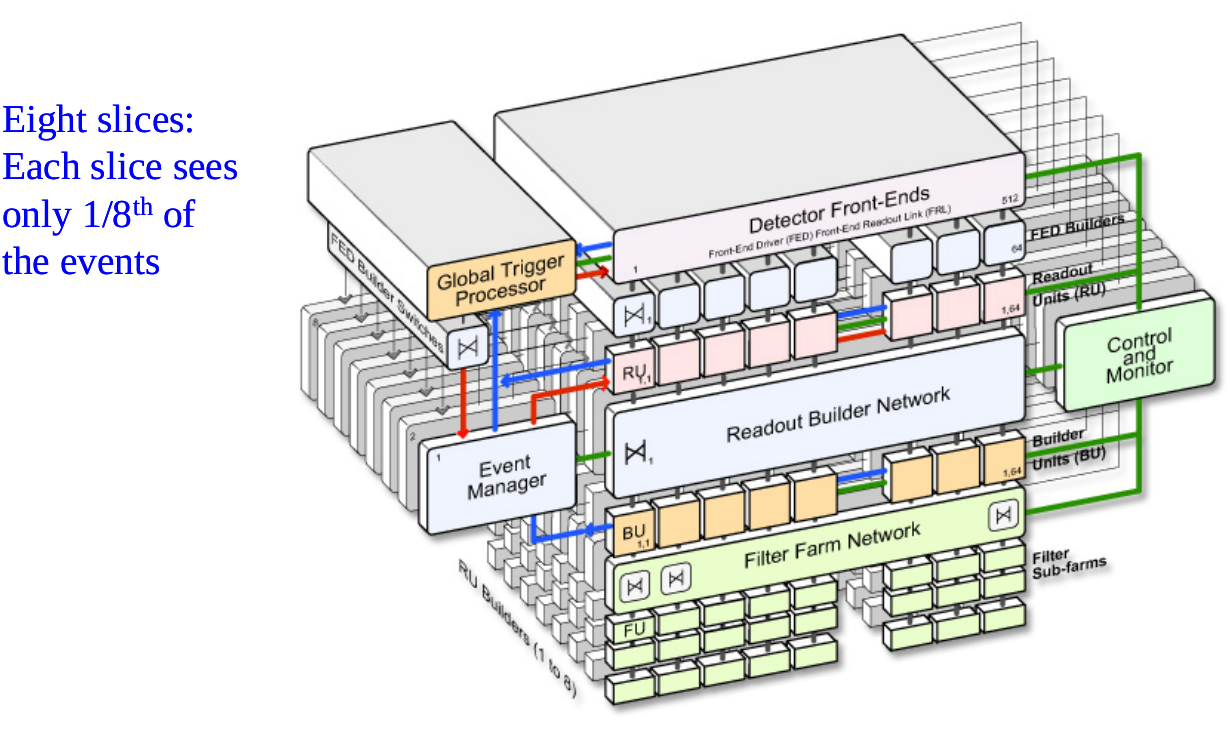}
\caption[]{The CMS slicing concept}
\label{fig:f20}
\end{figure}

In CMS the event building is factorized into a number of `slices', each
of which sees only a fraction of the total rate (see
\Figure~\ref{fig:f20}). This still requires a large total network
bandwidth (which has implications for the cost), but it avoids the
need for a very big central network switch. An additional advantage of
this approach is that the size of the system can be scaled, starting
with a few slices and adding more later (\eg as additional funding
becomes available).

\begin{figure}[ht]
\centering\includegraphics[width=.65\linewidth]{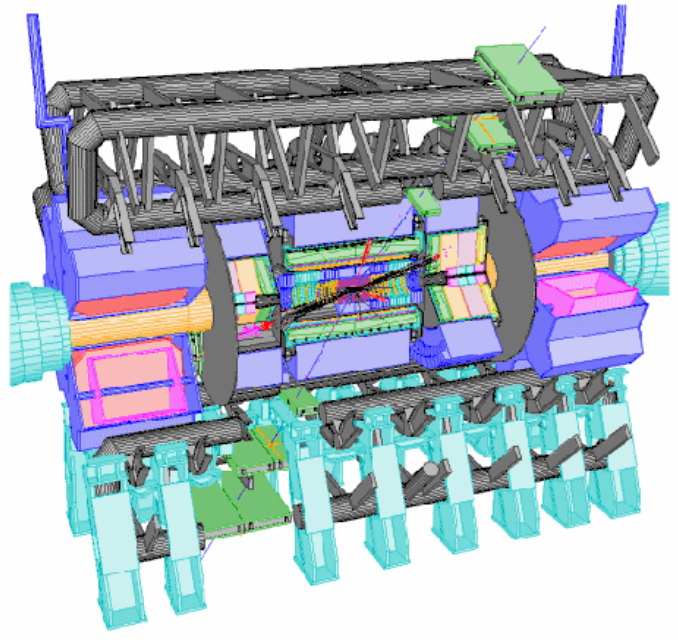}
\caption[]{The ATLAS region-of-interest concept\,---\,example of a
           dimuon event (see text)}
\label{fig:f21}
\end{figure}

In ATLAS the amount of data to be moved is reduced by using the
region-of-interest (RoI) mechanism (see \Figure~\ref{fig:f21}). Here,
the LVL1 trigger indicates the geographical location in the detector
of candidate objects. LVL2 then only needs to access data from the
RoIs, a small fraction of the total, even for the
calorimeter and muon detectors that participated in the LVL1
selection. This requires relatively complicated mechanisms to serve
the data selectively to the LVL2 trigger processors. 

In the example
shown in \Figure~\ref{fig:f21}, two muons are identified by LVL1. It can be seen
that only a small fraction of the detector has to be accessed to
validate the muons. In a first step only the data from the muon
detectors are accessed and processed, and many events will be rejected
where the more detailed analysis does not confirm the comparatively
crude LVL1 selection (\eg sharper {\pT} cut). For those events that
remain, the inner-tracker data will be accessed within the RoIs,
allowing further rejection (\eg of muons from decays in flight of
charged pions and kaons). In a last step, calorimeter information may
be accessed within the RoIs to select isolated muons (\eg to reduce
the high rate of events with muons from bottom and charm decays, while
retaining those from {\PW} and {\PZ} decays).

Concerning hardware implementation, the computer industry is putting
on the market technologies that can be used to build much of the
HLT/DAQ systems at the LHC. Computer network products now offer high
performance at affordable cost. Personal computers (PCs) provide
exceptional value for money in processing power, with high-speed
network interfaces as standard items. Nevertheless, custom hardware is
needed in the parts of the system that see the full LVL1 trigger
output rate (\textsim\,100\UkHz). This concerns the readout systems
that receive the detector data following a positive LVL1 trigger
decision, and (in ATLAS) the interface to the LVL1 trigger that
receives the RoI pointers. Of course, this is in
addition to the specialized front-end electronics associated with the
detectors that was discussed earlier (digitization, pipelines,
derandomizers, \etc).


\section{Physics requirements\,---\,two examples}

In the following, the physics requirements on the T/DAQ systems at LEP
and at the LHC are examined. These are complementary cases\,---\,at LEP
precision physics was the main emphasis, at the LHC discovery physics
will be the main issue. Precision physics at LEP needed accurate
determination of the absolute cross-section (\eg in the determination
of the number of light-neutrino species). Discovery physics at the LHC
will require sensitivity to a huge range of predicted processes with
diverse signatures (with very low signal rates expected in some
cases), aiming to be as sensitive as possible to new physics that has
not been predicted (by using inclusive signatures). This has to be
achieved in the presence of an enormous rate of Standard Model physics
backgrounds (the rate of proton--proton collisions at the LHC will be
$\OfOrd{10^9}$\UHz\,---\, $\sigma$\,\textsim\,100\Umb,
\Lumi{\text{\textsim}\,10^{34}}).

\subsection{Physics requirements at LEP}

Triggers at LEP aimed to identify all events coming from {\Pep\Pem}
annihilations with visible final states. At LEP-I, operating with
$\sqrt{s} \sim m_{\PZ}$, this included $\PZ \rightarrow \text{hadrons}$, 
$\PZ \rightarrow \Pep \Pem$, 
$\PZ \rightarrow \PGmp \PGmm$, and 
$\PZ \rightarrow \PGtp \PGtm$; 
at LEP-II, operating above the WW threshold, this included WW,
ZZ and single-boson events. Sensitivity was required even in cases
where there was little visible energy, \eg in the Standard Model for
$\Pep\Pem \rightarrow \PZ \PGg$,
with $\PZ \rightarrow \PGn \PGn$,
and in new-particle searches such as 
$\Pep\Pem \rightarrow \PSGcp \PSGcm$
for the case of small $\PSGcpm-\PSGcz$ mass difference that gives only
low-energy visible particles ($\PSGcz$ is the lightest supersymmetric
particle). In addition, the triggers had to retain some fraction of
two-photon collision events (used for QCD studies), and identify
Bhabha scatters (needed for precise luminosity determination).

The triggers could retain events with any significant activity in the
detector. Even when running at the {\PZ} peak, the rate of {\PZ}
decays was only \OfOrd{1}\UHz\,---\,physics rate was not an issue. The
challenge was in maximizing the efficiency (and acceptance) of the
trigger, and making sure that the small inefficiencies were very well
understood. The determination of absolute cross-section depends on
knowing the integrated luminosity and the experimental efficiency to
select the process in question (\ie the efficiency to trigger on the
specific physics process). Precise determination of the integrated
luminosity required excellent understanding of the trigger efficiency
for Bhabha-scattering events (luminosity determined from the rate of
Bhabha scatters within a given angular range). A major achievement at
LEP was to reach `per mil' precision.

The trigger rates (events per second) and the DAQ rates (bytes per
second) at LEP were modest as discussed in Section~\ref{sec:designlep}.

\subsection{Physics requirements at the LHC}

Triggers in the general-purpose proton--proton experiments at the LHC
(ATLAS~\cite{bib_ATLAS,bib_ATLAS2} and CMS~\cite{bib_CMS,bib_CMS2}) will have to retain as high as
possible a fraction of the events of interest for the diverse physics
programmes of these experiments. Higgs searches in and beyond the
Standard Model will include looking for 
$\PH \rightarrow \PZ \PZ \rightarrow \text{leptons}$
and also
$\PH \rightarrow \PQb \PAQb$.
Supersymmetry (SUSY) searches will be performed with and without the
assumption of R-parity conservation. One will search for other new
physics using inclusive triggers that one hopes will be sensitive to
unpredicted processes. In parallel with the searches for new physics,
the LHC experiments aim to do precision physics, such as measuring the
{\PW} mass and some B-physics studies, especially in the early phases of
LHC running when the luminosity is expected to be comparatively low.

In contrast to the experiments at LEP, the LHC trigger systems have a
hard job to reduce the physics event rate to a manageable level for
data recording and offline analysis. As discussed above, the design
luminosity \Lumi{\text{\textsim}\,10^{34}}, together with
$\sigma$\,\textsim\,100\Umb, implies an \OfOrd{10^9}\UHz{} interaction
rate. Even the rate of events containing leptonic decays of {\PW} and
{\PZ} bosons is \OfOrd{100}\UHz. Furthermore, the size of the events
is very large, \OfOrd{1}~Mbyte, reflecting the huge number of detector
channels and the high particle multiplicity in each event. Recording
and subsequently processing offline \OfOrd{100}\UHz~event rate per
experiment with an \OfOrd{1}~Mbyte event size is considered feasible, but
it implies major computing resources~\cite{bib_computing}. Hence, only a tiny
fraction of proton--proton collisions can be selected\,---\,taking the
order-of-magnitude numbers given above, the maximum fraction of
interactions that can be selected is \OfOrd{10^{-7}}. Note that the
general-purpose LHC experiments have to balance the needs of
maximizing physics coverage and reaching acceptable (\ie affordable)
recording rates.

The LHCb experiment~\cite{bib_LHCb}, which is dedicated to studying
B-physics, faces similar challenges to ATLAS and CMS. It will operate
at a comparatively low luminosity (\Lumi{\text{\textsim}\,10^{32}}),
giving an overall proton--proton interaction rate of
\textsim\,20\UMHz\,---\,chosen to maximize the rate of
single-interaction bunch crossings. The event size will be
comparatively small (\textsim\,100~kbytes) as a result of having fewer
detector channels and of the lower occupancy of the detector (due to
the lower luminosity with less pile-up). However, there will be a very
high rate of beauty production in LHCb\,---\,taking
$\sigma$\,\textsim\,~500~\textmu{}b, the production rate will be
\textsim\,100\UkHz\,---\,and the trigger must search for specific
B-decay modes that are of interest for physics analysis, with the aim
of recording an event rate of only \textsim\,200\UHz.

The heavy-ion experiment ALICE~\cite{bib_ALICE} is also very demanding,
particularly from the DAQ point of view. The total interaction rate
will be much smaller than in the proton--proton
experiments\,---\,\Lumi{\text{\textsim}\,10^{27}} is predicted to give a
rate \textsim\,8000\UHz{}
for Pb--Pb collisions. However, the event size will be huge due to the
high final-state multiplicity in Pb--Pb interactions at LHC energy. Up
to \OfOrd{10^4} charged particles will be produced in the central
region, giving an event size of up to \textsim\,40~Mbytes when the full
detector is read out. The ALICE trigger will select `minimum-bias' and
`central' events (rates scaled down to a total of about 40\UHz), and
events with dileptons (\textsim\,1\UkHz{} with only part of the detector
read out). Even compared to the other LHC experiments, the volume of
data to be stored and subsequently processed offline will be massive,
with a data rate to storage of \textsim\,1~Gbytes/s (considered to be
about the maximum affordable rate).

\section{Signatures of different types of particle}

\begin{figure}[ht]
\centering\includegraphics[width=.8\linewidth]{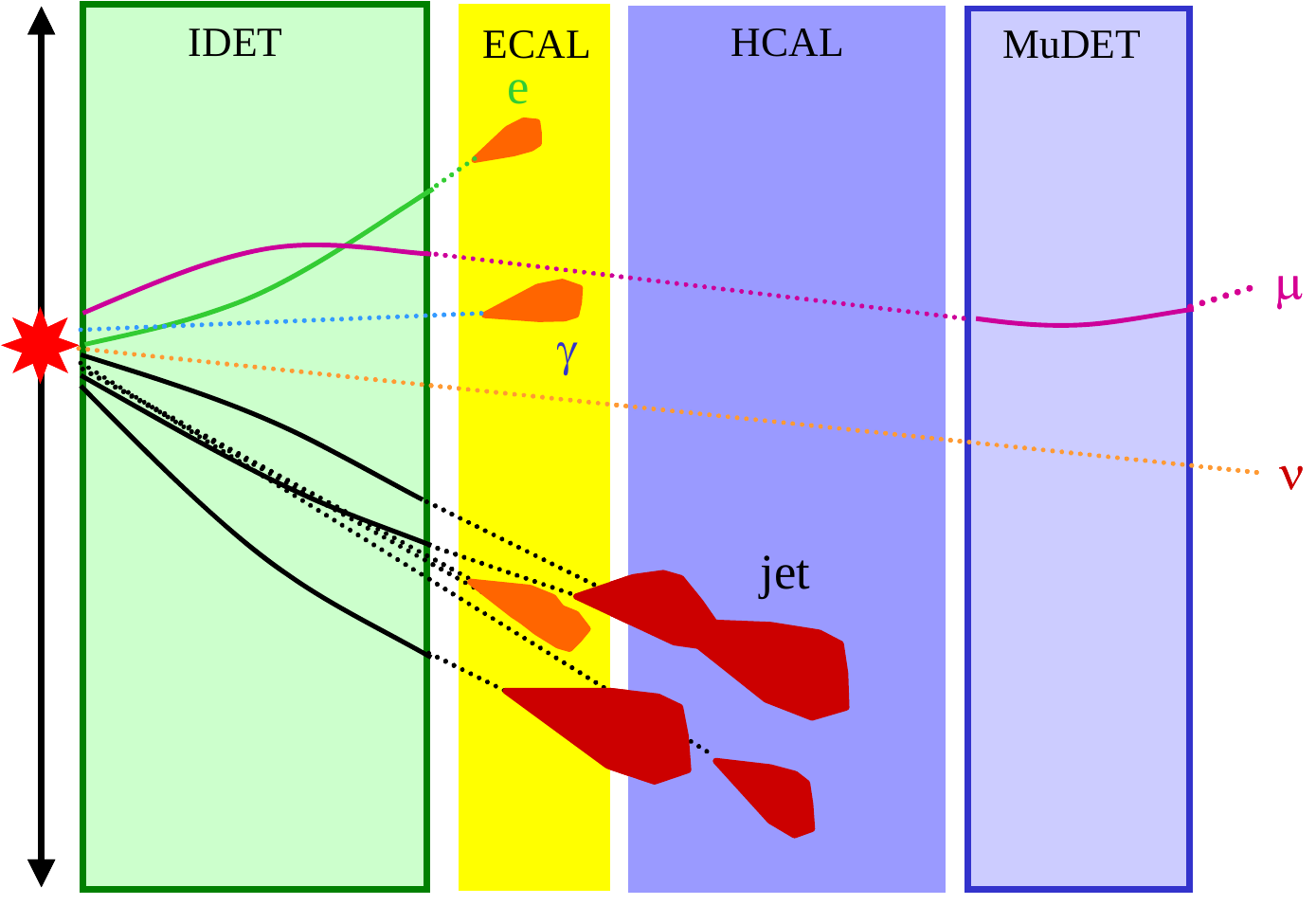}
\caption[]{Signatures of different types of particle in a generic detector}
\label{fig:f10}
\end{figure}

The generic signatures for different types of particle are
illustrated in \Figure~\ref{fig:f10}. Moving away from the interaction point (shown as a
star on the left-hand side of Fig.~\ref{fig:f10}), one finds the inner tracking
detector (IDET), the electromagnetic calorimeter (ECAL), the hadronic
calorimeter (HCAL) and the muon detectors (MuDET). Charged particles
(electrons, muons and charged hadrons) leave tracks in the
IDET. Electrons and photons shower in the ECAL, giving localized
clusters of energy without activity in the HCAL. Hadrons produce
larger showers that may start in the ECAL but extend into the
HCAL. Muons traverse the calorimeters with minimal energy loss and are
detected in the MuDET.

The momenta of charged particles are measured from the radii of
curvature of their tracks in the IDET which is embedded in a magnetic
field. A further measurement of the momenta of muons may be made in
the MuDET using a second magnet system. The energies of electrons,
photons and hadrons are measured in the calorimeters. Although
neutrinos leave the detector system without interaction, one can infer
their presence from the momentum imbalance in the event (sometimes
referred to as `missing energy'). Hadronic jets contain a mixture of
particles, including neutral pions that decay almost immediately into
photon pairs that are then detected in the ECAL. The jets appear as
broad clusters of energy in the calorimeters where the individual
particles will sometimes not be resolved.

\section{Selection criteria and trigger implementations at LEP}

The details of the selection criteria and trigger implementations at
LEP varied from experiment to
experiment~\cite{bib_ALEPH,bib_DELPHI,bib_L3,bib_OPAL}. Discussion of the example of
ALEPH is continued with the aim of giving a reasonably in-depth view
of one system. For triggering purposes, the detector was divided into
segments with a total of 60 regions in $\theta, \phi$ ($\theta$ is
polar angle and $\phi$ is azimuth with respect to the beam axis).
Within these segments, the following trigger objects were identified:

\begin{enumerate}
\item
 muon\,---\,requiring a track penetrating the hadron calorimeter and
  seen in the inner tracker;
\item
  charged electromagnetic (EM) energy\,---\,requiring an EM
  calorimeter cluster and a track in the inner tracker;
\item
  neutral EM energy\,---\,requiring an EM calorimeter cluster (with
  higher thresholds than in (2) to limit the rate to acceptable
  levels).
\end{enumerate}

In addition to the above local triggers, there were total-energy
triggers (applying thresholds on energies summed over large
regions\,---\,the barrel or a full endcap), a back-to-back tracks
trigger, and triggers for Bhabha scattering (luminosity monitor).

The LVL1 triggers were implemented using a combination of analog and
digital electronics. The calorimeter triggers were implemented using
analog electronics to sum signals before applying thresholds on the
sums. The LVL1 tracking trigger looked for patterns of hits in the
inner-tracking chamber (ITC) consistent with a track with $\pT > 1~
\mathrm{GeV}$~\footnote{Here, {\pT} is transverse momentum (measured
  with respect to the beam axis); similarly, {\ET} is transverse energy.}%
\,---\,at LVL2 the Time Projection Chamber (TPC) was used instead. The final decision was made
by combining digital information from calorimeter and tracking
triggers, making local combinations within segments of the detector,
and then making a global combination (logical OR of conditions).

\section{Selection criteria at LHC}

Features that distinguish new physics from the bulk of the
cross-section for Standard Model processes at hadron colliders are
generally the presence of high-{\pT} particles (or jets). For example,
these may be the products of the decays of new heavy particles. In
contrast, most of the particles produced in minimum-bias interactions
are soft (\pT\,\textsim\,1\UGeV{} or less). More specific signatures are
the presence of high-{\pT} leptons (\Pe, \PGm, \PGt), photons and/or
neutrinos. For example, these may be the products (directly or
indirectly) of new heavy particles. Charged leptons, photons and
neutrinos give a particularly clean signature (c.f.~low-{\pT} hadrons
in minimum-bias events), especially if they are `isolated' (\ie not
inside jets). The presence of heavy particles such as {\PW} and {\PZ}
bosons can be another signature for new physics\,---\,\eg they may be
produced in Higgs decays. Leptonic {\PW} and {\PZ} decays give a very
clean signature that can be used in the trigger. Of course it is
interesting to study {\PW} and {\PZ} boson production $per se$, and such
events can be very useful for detector studies (\eg calibration of the
EM calorimeters).

In view of the above, LVL1 triggers at hadron colliders search for the
following signatures (see \Figure~\ref{fig:f10}).
\begin{itemize}
\item 
High-{\pT} muons\,---\,these can be identified as charged particles that
penetrate beyond the calorimeters; a {\pT} cut is needed to control the
rate of muons from $\PGppm \rightarrow \PGmpm \PGn$ and
$\PKpm \rightarrow \PGmpm \PGn$ decays in flight, as well as 
those from semi-muonic beauty and charm decays.

\item
High-{\pT} photons\,---\,these can be identified as narrow clusters in
the EM calorimeter; cuts are made on transverse energy ($\ET >
\text{threshold}$), and isolation and associated hadronic transverse
energy ($\ET < \text{threshold}$), to reduce the rate due to
misidentified high-{\pT} jets.

\item
High-{\pT} electrons\,---\,identified in a similar way to photons,
although some experiments require a matching track as early as LVL1.

\item
High-{\pT} taus\,---\,identified as narrow clusters in the calorimeters
(EM and hadronic energy combined).

\item
High-{\pT} jets\,---\,identified as wider clusters in the calorimeters
(EM and hadronic energy combined); note that one needs to cut at very high
{\pT} to get acceptable rates given that jets are the dominant high-{\pT}
process.

\item
Large missing {\ET} or scalar {\ET}.

\end{itemize}

Some experiments also search for tracks from displaced secondary
vertices at an early stage in the trigger selection.

The trigger selection criteria are typically expressed as a list of
conditions that should be satisfied\,---\,if any of the conditions is
met, a trigger is generated (subject to dead-time requirements,
\etc). In these notes, the list of conditions is referred to as the 
`trigger menu', although the name varies from experiment to
experiment. An illustrative example of a LVL1 trigger menu for
high-luminosity running at LHC includes the following (rates \cite{bib_ATLAS} are
given for the case of ATLAS at  \Lumi{\text{\textsim}\,10^{34}}):
\begin{itemize}
\item one or more muons with $\pT > 20\UGeV$ (rate
      \textsim\,11\UkHz);
\item two or more muons each with $\pT > 6\UGeV$ (rate 
      \textsim\,1\UkHz);
\item one or more \Pe/\PGg{} with $\ET > 30\UGeV$ (rate 
      \textsim\,22\UkHz);
\item two or more \Pe/\PGg{} each with $\ET > 20\UGeV$ (rate
      \textsim\,5\UkHz);
\item one or more jets with $\ET > 290\UGeV$ (rate 
      \textsim\,200\UHz);
\item one or more jets with $\ET > 100\UGeV$ and missing-$\ET > 100\UGeV$ 
      (rate \textsim\,500\UHz);
\item three or more jets with $\ET > 130\UGeV$ (rate 
      \textsim\,200\UHz);
\item four or more jets with $\ET > 90\UGeV$ (rate \textsim\,200\UHz).
\end{itemize}

The above list represents an extract from a LVL1 trigger menu,
indicating some of the most important trigger requirements\,---\,the
full menu would include many items in addition (typically more than~100 items
in total). The additional items are expected to include the following:

\begin{itemize}
\item $\PGt$ (or isolated single-hadron) candidates;

\item combinations of objects of different types (\eg muon \emph{and}
      \Pe/\PGg);

\item pre-scaled\footnote{Some triggers may be `pre-scaled'\,---\,this
        means that only every \emph{N}\textsuperscript{th} event satisfying
        the relevant criteria is recorded, where \emph{N} is a parameter
        called the pre-scale factor; this is useful for collecting
        samples of high-rate triggers without swamping the T/DAQ
        system.}  
      triggers with lower thresholds;

\item triggers needed for technical studies and to aid understanding
      of the data from the main triggers (\eg trigger on bunch
      crossings at random to collect an unbiased data sample).
\end{itemize}

As for the LVL1 trigger, the HLT has a trigger menu that describes
which events should be selected. This is illustrated in
\Table~\ref{tab:triggerrates} for the example of CMS, assuming a
luminosity for early running of \Lumi{\text{\textsim}\,10^{33}}. The
total rate of \textsim\,100\UHz{} contains a large fraction of events
that are useful for physics analysis. Lower thresholds would be
desirable, but the physics coverage has to be balanced against
considerations of the offline computing cost. Note that there are
large uncertainties on the rate calculations.

\begin{table}[h]
\caption[]{Estimated high-level trigger rates for
           \Lumi{\text{\textsim}2\times 10^{33}} (CMS numbers from
           Ref.~\cite{bib_CMS})}
\label{tab:triggerrates}

\renewcommand{\arraystretch}{1.1}%
\centering
\begin{tabularx}{\linewidth}{@{}X@{\quad}r@{}}
\hline\hline
\textbf{Trigger configuration}
   & \textbf{Rate}\quad \\\hline
One or more electrons with $\pT > 29\UGeV$, or two or more electrons with
$\pT > 17\UGeV$ 
   & \textsim\,34\UHz \\
One or more photons with $\pT > 80\UGeV$, or two or more photons with 
$\pT > 40, 25\UGeV$
   & \textsim\,9\UHz  \\
One or more muons with $\pT > 19\UGeV$, or two or more muons with 
$\pT > 7\UGeV$
   & \textsim\,29\UHz \\
One or more taus with $\pT > 86\UGeV$, or two or more taus with 
$\pT > 59\UGeV$
   & \textsim\,4\UHz  \\
One or more jets with $\pT > 180\UGeV$ \emph{and} 
missing-\ET $ > 123\UGeV$
   & \textsim\,5\UHz  \\
One or more jets with $\pT > 657\UGeV$, or three or more jets with 
$\pT > 247\UGeV$, or four or more jets with $\pT > 113\UGeV$
   & \textsim\,9\UHz  \\
Others (electron and jet, b-jets, \etc)
   & \textsim\,7\UHz  \\
\hline\hline
\end{tabularx}
\end{table}

\newpage
A major challenge lies in the HLT/DAQ software. The event-selection
algorithms for the HLT can be subdivided, at least logically, into
LVL2 and LVL3 trigger stages. These might be performed by two separate
processor systems (\eg ATLAS), or in two distinct processing steps
within the same processor system (\eg CMS). The algorithms have to be
supported by a software framework that manages the flow of data,
supervising an event from when it arrives at the HLT/DAQ system until
it is either rejected, or accepted and recorded on permanent
storage. This includes software for efficient transfer of data to the
algorithms. In addition to the above, there is a large amount of
associated online software (run control, databases, book-keeping,
\etc).

\section{LVL1 trigger design for the LHC}

A number of design goals must be kept in mind for the LVL1 triggers at
the LHC. It is essential to achieve a very large reduction in the physics
rate, otherwise the HLT/DAQ system will be swamped and the dead-time
will become unacceptable. In practice, the interaction rate,
\OfOrd{10^9}\UHz, must be reduced to less than 100\UkHz{} in ATLAS and
CMS. Complex algorithms are needed to reject the background while
keeping the signal events.

Another important constraint is to achieve a short
latency\,---\,information from all detector elements
(\OfOrd{10^7 \text{--} 10^8} channels!) has to be held on the
detector pending the LVL1 decision. The pipeline memories that do this
are typically implemented in ASICs (application-specific integrated
circuits), and memory size contributes to the cost. Typical LVL1
latency values are a few microseconds (\eg less than 2.5 {\textmu}s in
ATLAS and less than 3.2 {\textmu}s in CMS).

A third requirement is to have flexibility to react to changing
conditions (\eg a wide range of luminosities)
and\,---\,it is hoped\,---\,to new physics! The algorithms must be
programmable, at least at the level of parameters (thresholds, \etc).

\subsection{Case study\,---\,ATLAS \Pe/\PGg{} trigger}

\begin{figure}[p]
\centering\includegraphics[width=.65\linewidth]{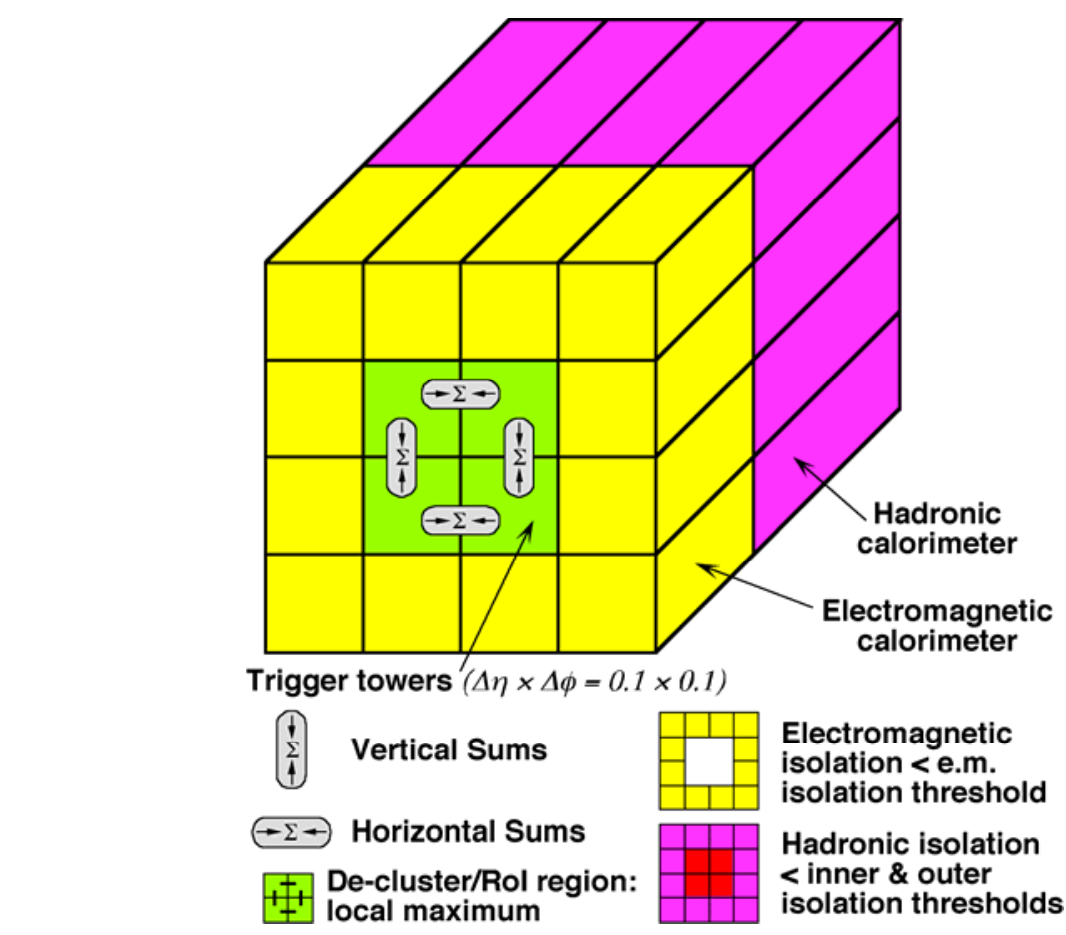}
\caption[]{ATLAS {\Pe/\PGg} trigger algorithm}
\label{fig:f16}

\bigskip
\centering\includegraphics[bb=30 40 345 400, clip, width=.7\linewidth]{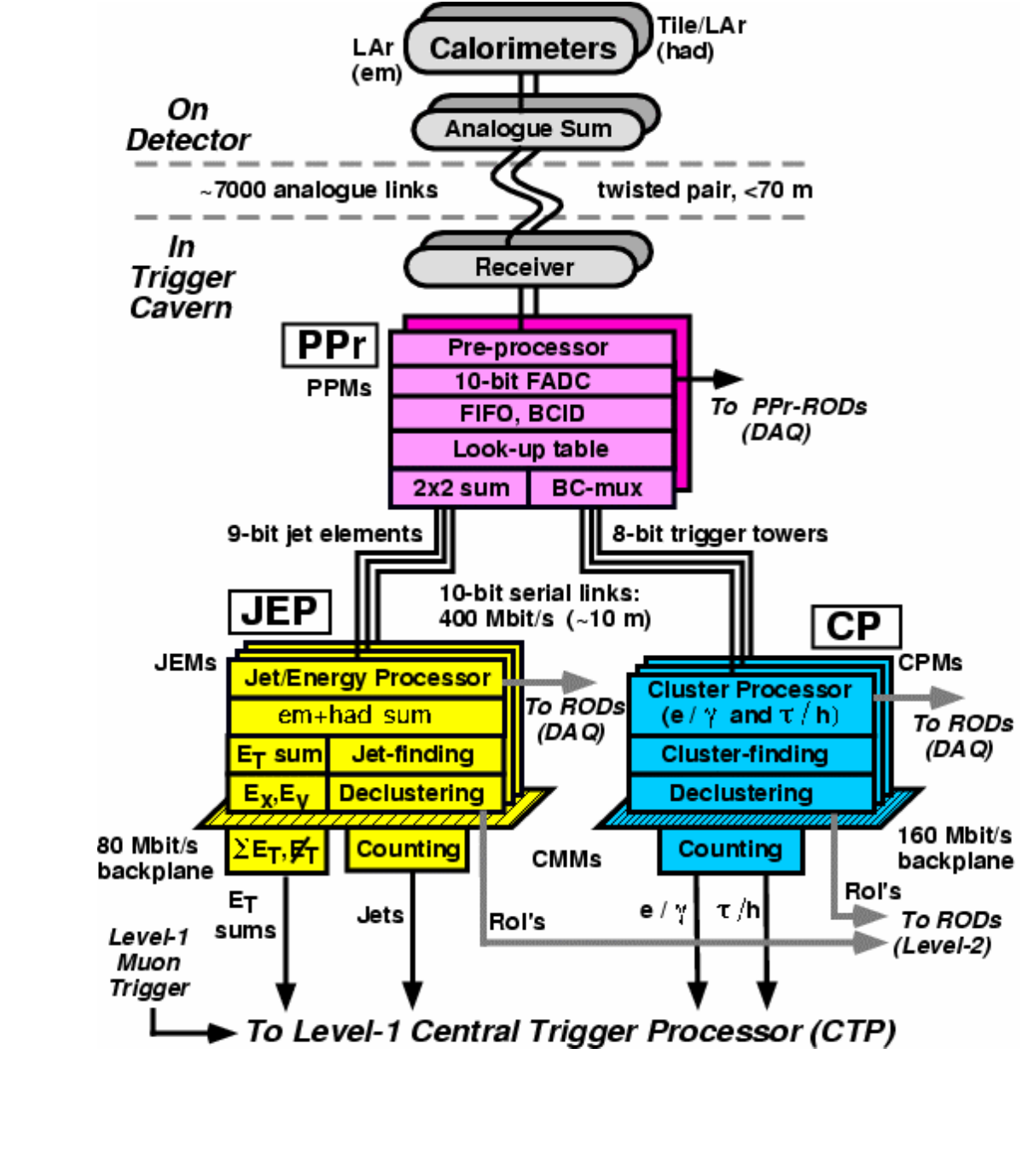}
\caption[]{Overview of the ATLAS LVL1 calorimeter trigger}
\label{fig:f17}

\end{figure}

The ATLAS \Pe/\PGg{} trigger algorithm can be used to illustrate the
techniques used in LVL1 trigger systems at LHC. It is based on $4
\times 4$ `overlapping, sliding windows' of trigger towers as
illustrated in \Figure~\ref{fig:f16}. Each trigger tower has a lateral
extent of $0.1 \times 0.1$ in $\eta,\phi$ space, where $\eta$ is
pseudorapidity and $\phi$ is azimuth. There are about 3500 such towers
in each of the EM and hadronic calorimeters. Note that each tower
participates in calculations for 16 windows. The algorithm requires a
local maximum in the EM calorimeter to define the $\eta\text{--}\phi$
position of the cluster and to avoid double counting of extended
clusters (so-called `declustering'). It can also require that the
cluster be isolated, \ie little energy surrounding the cluster in the EM
calorimeter or the hadronic calorimeter.

The implementation of the ATLAS LVL1 calorimeter trigger \cite{bib_ATLASL1calo} is sketched
in \Figure~\ref{fig:f17}. Analog electronics on the detector sums
signals from individual calorimeter cells to form trigger-tower
signals. After transmission to the `pre-processor' (PPr), which is
located in an underground room close to the detector and shielded against radiation, the tower signals are received and
digitized; then the digital data are processed to obtain estimates of
{\ET} per trigger tower for each BC. At this point in the processing
chain (\ie at the output of the PPr), there is an `$\eta\text{--}\phi$
matrix' of the {\ET} per tower in each of the EM and hadronic
calorimeters that gets updated every 25~ns.

The tower data from the PPr are transmitted to the cluster processor
(CP). Note that the CP is implemented with very dense electronics so
that there are only four crates in total. This minimizes the number of
towers that need to be transmitted (`fanned out') to more than one
crate. Fan out is required for towers that contribute to windows for
which the algorithmic processing is implemented in more than one
crate. Also, within each CP crate, trigger-tower data need to be
fanned out between electronic modules, and then between processing
elements within each module. Considerations of connectivity and
data-movement drive the design.

In parallel with the CP, a jet/energy processor (JEP) searches for jet
candidates and calculates missing-{\ET} and scalar-{\ET} sums. This is
not described further here.

\begin{figure}[ht]
\centering\includegraphics[width=.65\linewidth]{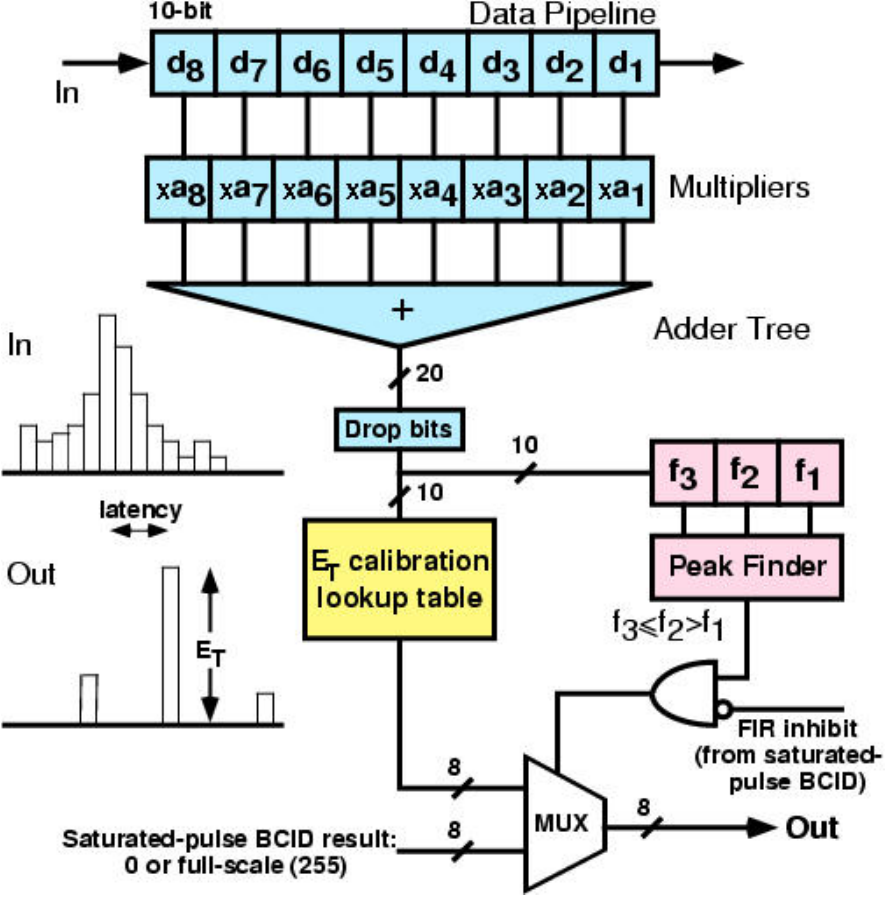}
\caption[]{Bunch-crossing identification}
\label{fig:f18}
\end{figure}

A very important consideration in designing the LVL1 trigger is the
need to identify uniquely the BC that produced the interaction of
interest. This is not trivial, especially given that the calorimeter
signals extend over many BCs. In order to assign observed energy
deposits to a given BC, information has to be combined from a sequence
of measurements. \Figure[b]~\ref{fig:f18} illustrates how this is done
within the PPr (the logic is repeated \textsim\,7000 times so that
this is done in parallel for all towers). The raw data for a given
tower move along a pipeline that is clocked by the 40\UMHz{} BC
signal. The multipliers together with the adder tree implement a
finite-impulse-response filter whose output is passed to a peak finder
(a peak indicates that the energy was deposited in the BC currently
being examined) and to a look-up table that converts the peak
amplitude to an {\ET} value. Special care is taken to avoid BC
misidentification for very large pulses that may get distorted in the
analog electronics, since such signals could correspond to the most
interesting events. The functionality shown in~\Figure~\ref{fig:f18} is
implemented in ASICs (four channels per ASIC).

The transmission of the data (\ie the {\ET} matrices) from the PPr to
the CP is performed using a total of~5000 digital links each
operating at 400 Mbits/s (each link carries data from two towers using
a technique called BC multiplexing~\cite{bib_ATLASL1calo}). Where fan out is
required, the corresponding links are duplicated with the data being
sent to two different CP crates. Within each CP crate, data are shared
between neighbouring modules over a very high density crate back-plane
(\textsim\,800 pins per slot in a 9U crate; data rate of 160~Mbits/s
per signal pin using point-to-point connections). On each of the
modules, data are passed to eight large field-programmable gate arrays
(FPGAs) that perform the algorithmic processing, fanning out signals
to more than one FPGA where required.

As an exercise, it is suggested that students make an
order-of-magnitude estimate of the total bandwidth between the PPr and
the CP, considering what this corresponds to in terms of an equivalent
number of simultaneous telephone calls\footnote{One may assume an
  order-of-magnitude data rate for voice calls of 10 kbits/s\,---\,for
  example, the GSM mobile-phone standard uses a 9600 bit/s digital link
  to transmit the encoded voice signal.}.

The \Pe/\PGg{} (together with the \PGt/\Ph) algorithms are implemented
using FPGAs. This has only become feasible thanks to recent advances
in FPGA technology since very large and very fast devices are
needed. Each FPGA handles an area of $4 \times 2$ windows, requiring
data from $7 \times 5$ towers in each of the EM and hadronic
calorimeters. The algorithm is described in a programming language
(\eg VHDL) that can be converted into the FPGA configuration
file. This gives flexibility to adapt algorithms in the light of
experience\,---\,the FPGAs can be reconfigured \emph{in situ}. Note
that parameters of the algorithms can be changed easily and quickly,
\eg as the luminosity falls during the course of a coast of the beams
in the LHC machine, since they are held in registers inside the FPGAs
that can be modified at run time (\ie~there is no need to change the
`program' in the FPGA).

\section{High-level trigger algorithms}

There was not time in the lectures for a detailed discussion of the algorithms that are used in the HLT. However, it is useful to consider the case of the electron selection that follows after the first-level trigger. The LVL1 \Pe/\PGg{}  trigger is already very selective, so it is necessary to use complex algorithms and full-granularity, full-precision detector data in the HLT. 

A calorimeter selection is made applying a sharper \ET~cut (better resolution than at LVL1) and shower-shape variables that distinguish between the electromagnetic showers of an electron or photon on one hand, and activity from jets on the other hand. The shower-shape variables use both lateral and depth profile information. Then, for electrons, a requirement is made of an associated track in the inner detector, matching the calorimeter cluster in space, and with consistent momentum and energy measurements from the inner detector and calorimeter respectively.

\begin{figure}[ht]
\centering\includegraphics[width=.5\linewidth]{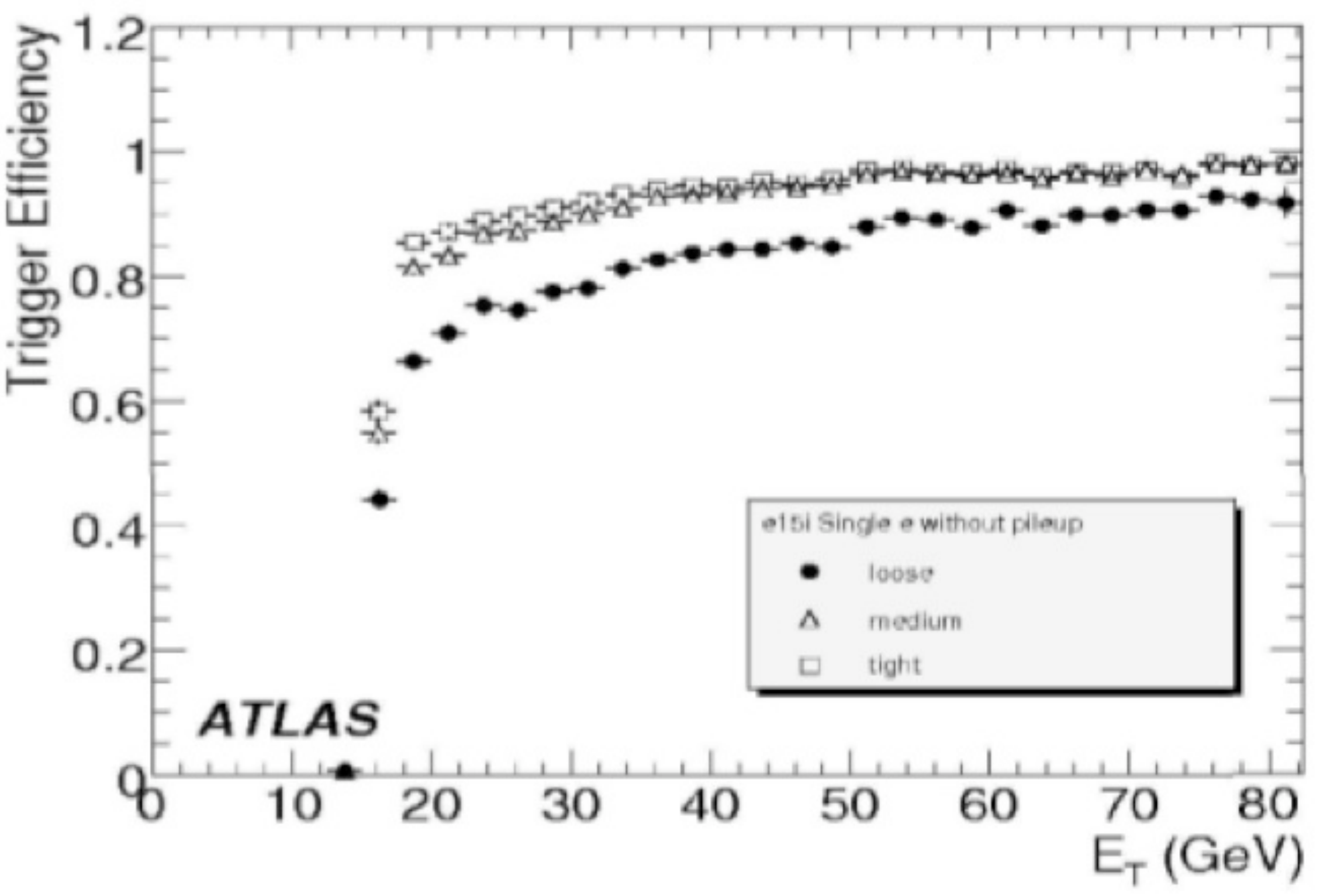}
\caption[]{Trigger efficiency versus electron \ET~for three different offline selections of the reference sample}
\label{fig:HLTe}
\end{figure}

Much work is going on to develop the algorithms and tune their many parameters to optimize their signal efficiency and background rejection. So far this has been done with simulated data, but further optimization will be required once samples of electrons are available from offline reconstruction of real data. It is worth noting that the efficiency value depends on the signal definition as shown in \Fref{fig:HLTe}, an example of a study taken from Ref.~\cite{bib_Navara}. Here the trigger efficiency is shown, as a function of electron transverse energy, relative to three different offline selections. With a loose offline selection, the trigger is comparatively inefficient, whereas it performs much better relative to the tighter offline cuts. This is related to the optimization of the trigger both for signal efficiency (where loose cuts are preferable) and for background rejection (where tighter cuts are required).

\section{Commissioning of the T/DAQ systems at LHC}

Much more detail on the general commissioning of the LHC experiments can be found in the lectures of Andreas Hoecker at this School~\cite{bib_Hoecker}. Here an attempt is made to describe how commissioning of the T/DAQ systems started in September 2008.

On 10 September 2008 the first beams passed around the LHC in both the clockwise and anti-clockwise directions, but with only one beam at a time (so there was no possibility of observing proton--proton collisions). The energy of the protons was 450 GeV which is the injection energy prior to acceleration; acceleration to higher energies was not attempted. 

As a first step, the beams were brought around the machine and stopped on collimators such as those upstream of the ATLAS experiment. Given the huge number of protons per bunch, as well as the sizeable beam energy, extremely large numbers of secondary particles were produced, including muons that traversed the experiment depositing energy in all of the detector systems.

Next, the collimators were removed and the beams were allowed to circulate around the machine for a few turns and, after some tuning, for a few tens of turns. Subsequently, the beams were captured by the radio-frequency system of the LHC and circulated for periods of tens of minutes. 

\begin{figure}[ht]
\centering\includegraphics[width=.8\linewidth]{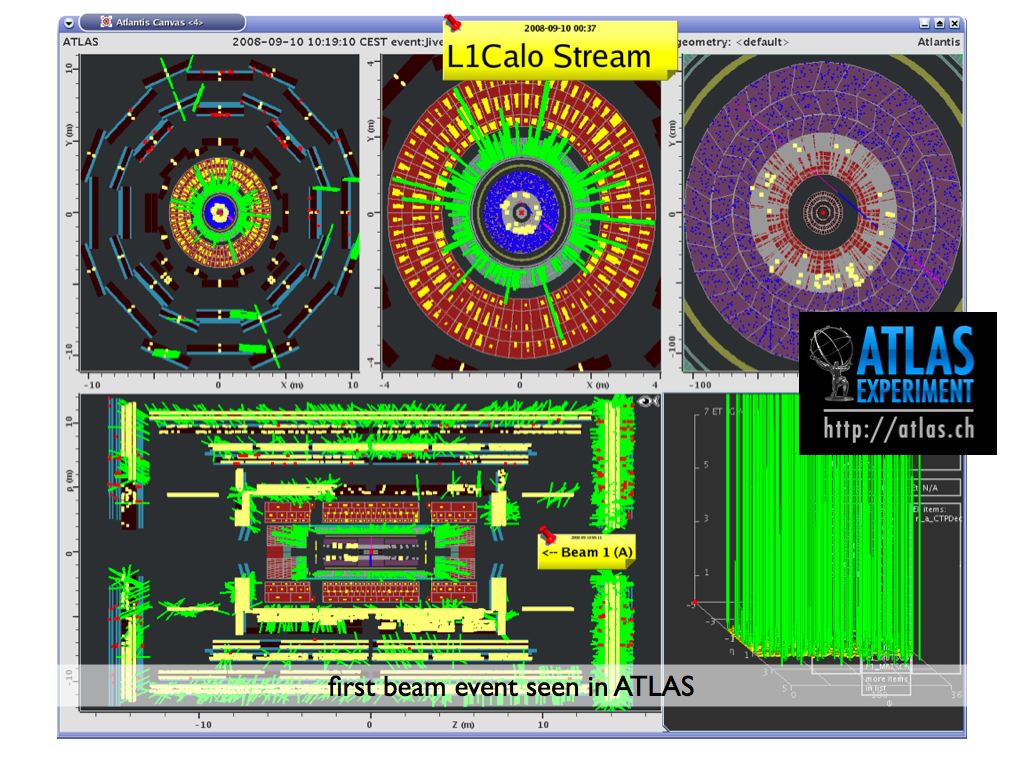}
\caption[]{The first beam-splash event in ATLAS}
\label{fig:FirstEv}
\end{figure}

The first day of LHC operations was very exciting for all the people working on the experiments. There was a very large amount of media interest, with television broadcasts from various control rooms around the CERN site. It was a particularly challenging time for those working on the T/DAQ systems who were anxious to see if the first beam-related events would be identified and recorded successfully. Much to the relief of the author, the online event display of ATLAS soon showed a spectacular beam-splash event produced when the beam particles hit the collimator upstream of the experiment. The first ATLAS event is shown in \Fref{fig:FirstEv}; similar events were seen by the other experiments. 

Analysis of the beam-spash events provided much useful information for commissioning the detectors and also the trigger. For example, the relative timing of different detector elements could be measured allowing the adjustment of programmable delays to the correct settings. The very large amount of activity in the events had the advantage that signals were seen in an unusually large fraction of the detector channels. 

\begin{figure}[ht]
\centering\includegraphics[width=.8\linewidth]{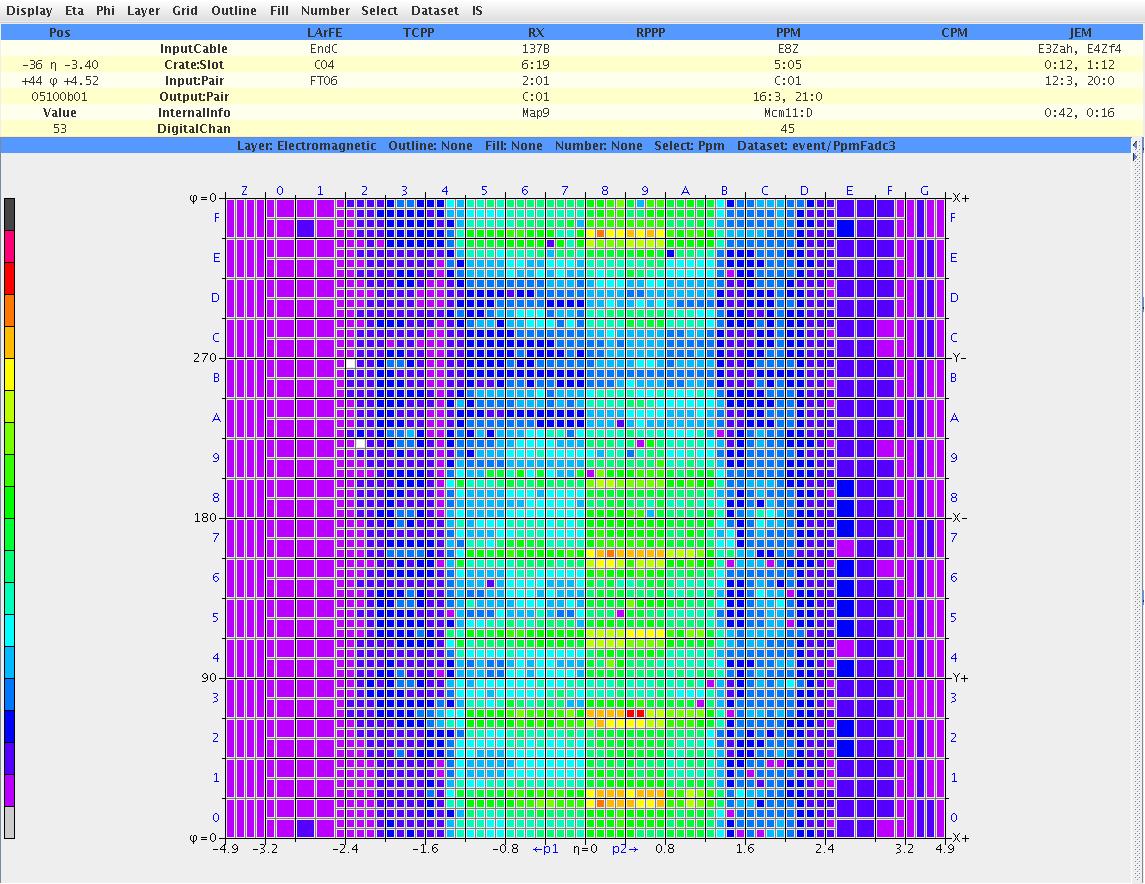}
\caption[]{LVL1 calorimeter trigger energy grid for a beam-splash event}
\label{fig:L1calo_splash}
\end{figure}

An example of a very early study done with beam-splash events in shown in  \Fref{fig:L1calo_splash} which plots \ET~versus $\eta$~and $\phi$~for the ATLAS LVL1 calorimeter trigger readout. The \ET~values are colour coded; $\eta$~is along the $x$-axis and $\phi$~is along the $y$-axis. The eight-fold $\phi$~structure of the ATLAS magnets can be seen, as well as the effects of the tunnel floor and heavy mechanical support structures that reduced the flux of particles reaching the calorimeters in the bottom part of the detector ($\phi~\approx$~270 degrees). The difference in absolute scale between the left-hand and right-hand sides of the plot is attributed to the fact that timing of the left-hand side was actually one bunch-crossing away from ideal when the data were collected; the timing calibration was subsequently adjusted as a result of these observations. 

At least in ATLAS, the first beam-spash events were recorded using triggers that had already been tested, with a free-runing 40~MHz clock, for cosmic-ray events. This approach was appropriate because of the importance of recording the first beam-related activity in the detector before the local beam instrumentation had been calibrated. However, it was crucial to move on as rapidly as possible to establish a precise and stable time reference.

Once beam-related activity had been seen in all of the LHC experiments, stopping the beam on the corresponding collimators, all of the collimators were removed and the beam was allowed to circulate. The first circulating beams passed around the LHC for only a short period of time, corresponding to a few turns initially, rising to a few tens of turns. For the 27~km LHC circumference, the orbit period is about 89~$\mu$s. 

\begin{figure}[ht]
\centering\includegraphics[width=.8\linewidth]{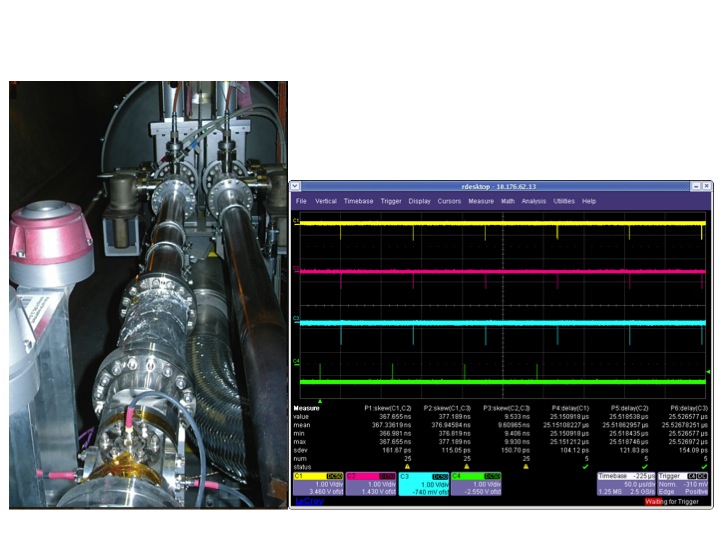}
\caption[]{Photograph of beam pick-up instrumentation (left) and display of timing signals recorded on a digital oscilloscope (right). The upper three traces are `orbit' signals from the LHC machine, whereas the bottom one is the (inverted) discriminated signal from the beam pick-up.}
\label{fig:BPTX}
\end{figure}

Upstream of the LHC detectors (and upstream of the collimators) are passive beam pick-ups that provide electrical signals induced by the passage of the proton beams. The photograph in the left-hand side of  \Fref{fig:BPTX} shows the beam pick-up for one of the beams in an LHC experiment. Three of the four cables that carry the signals can be seen. The analog signals from electrodes above, below, to the left and to the right of the beam are combined (analog sum). The resulting signal is fed to an oscilloscope directly and also via a discriminator (an electronic device that provides a logical output signal when the analog input signal exceeds a preset threshold, see Section~\ref{sec:design}).

On the right-hand side of \Fref{fig:BPTX} can be seen a plot, from CMS, of the relative timing of different signals. The upper three traces are `orbit' signals provided by the LHC machine, whereas the bottom trace is the discriminated beam pick-up signal. As can be seen, the pick-up signal is present for only four turns and then disappears. The reason for this is that after a few turns the protons de-bunched and the analog signal from the pick-ups became too small to fire the discriminator. Similar instrumentation and timing calibration studies were used in all of the LHC experiments.

A key feature of the beam pick-ups is that they provide a stable time reference with respect to which other signals can be aligned. The time of arrival of the beam pick-up signal, relative to the moment when the beam passes through the centre of the LHC detector, depends only on the proton time of flight from the beam pick-up position to the centre of the detector, propagation delays of the signal along the electrical cables, and the response time of the electronic circuits (which is very short).

Thanks to thorough preparations, the beam pick-up signals and their timing relative to the trigger could be measured as soon as beam was injected. Programmable delays could then be adjusted to align in time inputs to the trigger from the beam pick-ups and from other sources. For example, in ATLAS, the beam pick-up inputs were delayed so that they would have the same timing as other inputs that had already been adjusted using cosmic-rays.

Once the timing of the beam pick-up inputs to the trigger had been adjusted so as to initiate the detector readout for the appropriate bunch crossing (BC), i.e., to read out a time-frame that would contain the detector signals produced by beam-related activity, they could be used to provide the trigger for subsequent running.

It is worth noting that the steps described above to set up the timing of the trigger were completed within just a few hours on the morning of 10 September 2008. From then onwards the beam pick-ups represented a stable time reference with respect to which other elements in the trigger and in the detector readout systems could be adjusted. 

As already indicated, all of the beam operations in September 2008 were with just a single beam in the LHC. Operations were performed with beams circulating in both the clockwise and anti-clockwise directions. Beam activity in the detectors was produced by beam splash (beam stopped on collimators upstream of the detectors producing a massive number of secondary particles) or by beam-halo particles (produced when protons lost from the beam upstream of the detectors produced one or more high-momentum muons that traversed the detectors). In both cases one has to take into account the time of flight of the particles that reach one end of the detector before the other end. In contrast, beam--beam interactions have symmetric timing for the two ends of the detector.

The work on timing calibration performed over the days following the LHC start up can be illustrated by the case of ATLAS. Already on 10 September both sets of beam pick-ups had been commissioned (with beams circulating in the clockwise and anti-clockwise directions) giving a fixed time reference with respect to which the rest of the trigger, and indeed the rest of the experiment, could be aligned. 

\begin{figure}[ht]
\centering\includegraphics[width=.4\linewidth]{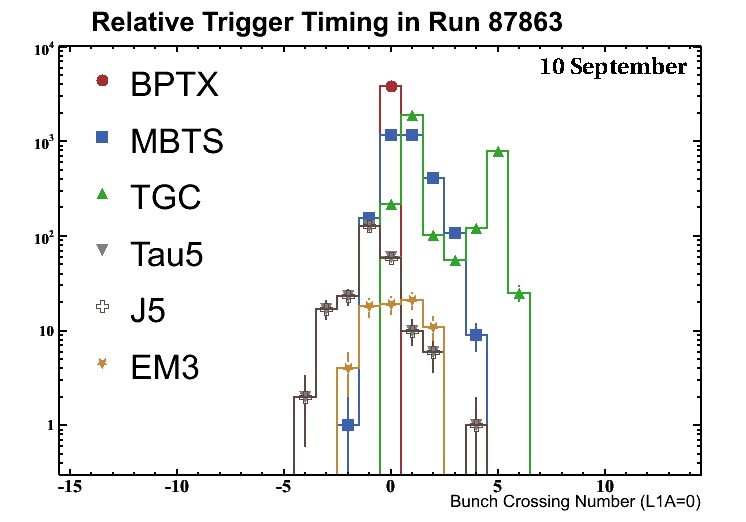}
\centering\includegraphics[width=.4\linewidth]{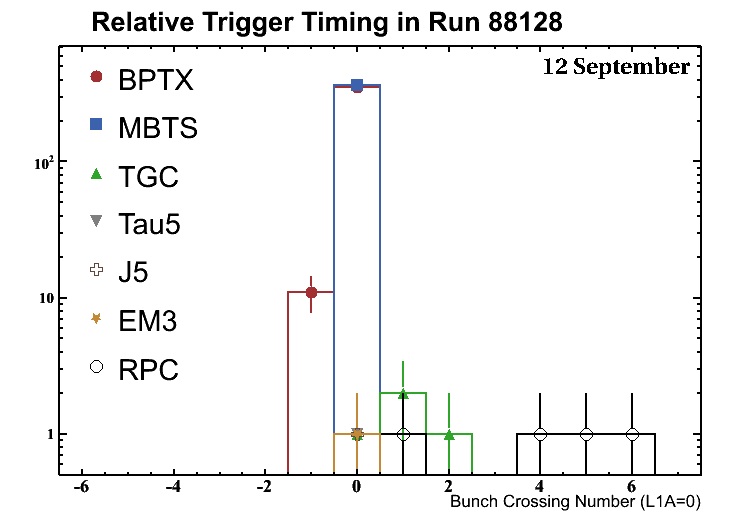}
\caption[]{Progress on timing-in ATLAS between 10 and 12 September 2008}
\label{fig:Timing_In}
\end{figure}

The situation on 10 September is summarized in the left-hand plot of \Fref{fig:Timing_In}. The beam pick-up signal, labelled `BPTX' in the figure, is the reference. The relative time of arrival of other inputs to the trigger is shown in units of BC number (i.e. one unit corresponds to 25~ns which is the nominal bunch-crossing interval at LHC). Although there is a peak at the nominal timing (bunch-number zero) in the distributions based on different trigger inputs --- the Minimum-Bias Trigger Scintillators (MBTS), the Thin-Gap Chamber (TGC) forward muon detectors, and the Tau5, J5 and EM3 items from the calorimeter trigger --- the distribution is broad.

Prompt analysis and interpretation of the data allowed the timing to be understood and calibration corrections to be applied. Issues addressed included programming delay circuits to correct for time of flight of the particles according to the direction of the circulating beam and tuning the relative timing of triggers from different parts of the detector or from different detector channels.

The situation two days later on 12 September is summarized in the right-hand plot of \Fref{fig:Timing_In}. It is important to note that the scale is logarithmic --- the vast majority of the triggers are aligned correctly in the nominal bunch crossing. Although shown in the plot, the input from the Resistive Plate Chambers (RPC) barrel muon detectors, which see very little beam-halo activity in single-beam operation, had not been timed-in.

As can be seen from the above, very significant progress was made on setting up the timing of the experiments within the first few days of single-beam operations at LHC. The experimental teams were eagerly awaiting further beam time and the first collisions that would have allowed them to continue the work. However, unfortunately, on 19 September there was a serious accident with the LHC machine that required a prolonged shutdown for repairs and improvements. Nevertheless, when the LHC restarts one will be able to build on the work that was already done (complemented by many further studies that were done using cosmic rays during the machine shutdown).

\begin{figure}[ht]
\centering\includegraphics[width=.5\linewidth]{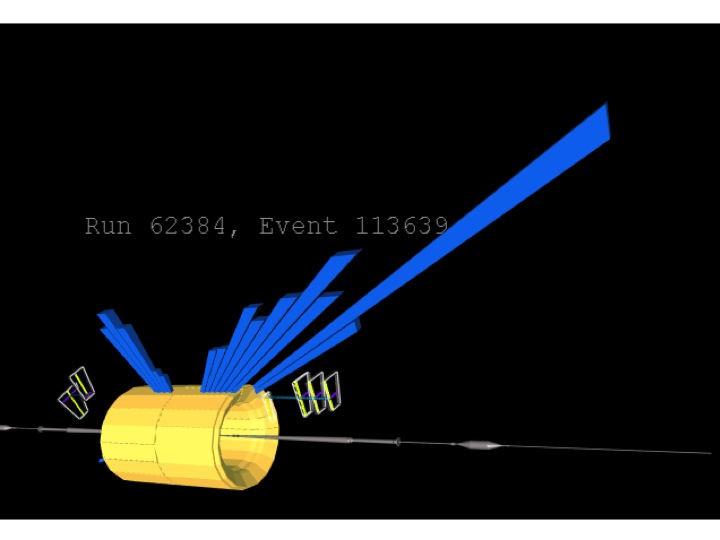}
\caption[]{A beam-halo event in CMS}
\label{fig:BeamHevent}
\end{figure}

A huge amount of work has been done using the beam-related data that were recorded in September 2008, as discussed in much more detail in the lectures of Andreas Hoecker at this School~\cite{bib_Hoecker}. A very important feature of these data is that activity is seen in the same event in several detector subsystems which allows one to check the relative timing and spatial alignment. Indeed the fact that the same event is seen in the different subdetectors is reassuring --- some previous experiments had teething problems where the readout of some of the subdetectors became desynchronized! A nice example of a beam-halo event recorded in CMS  is shown in  \Fref{fig:BeamHevent}. Activity can be seen in the Cathode-Strip Chamber (CSC) muon detectors at both ends of the experiment and also in the hadronic calorimeter.

\begin{figure}[ht]
\centering\includegraphics[width=.5\linewidth]{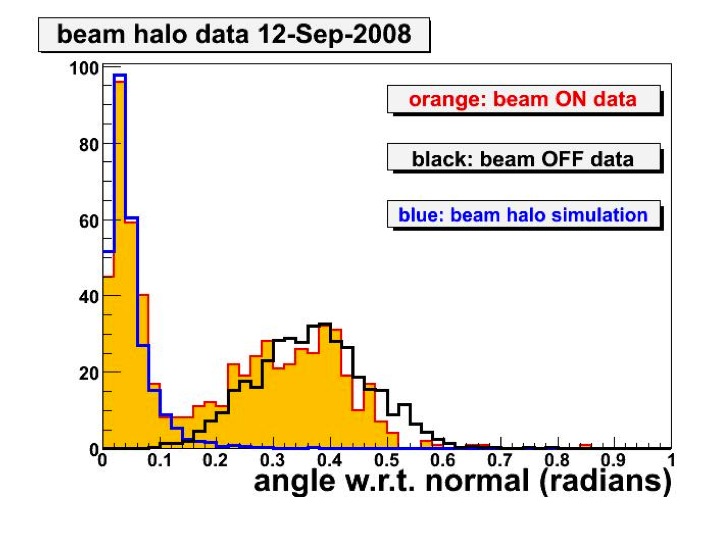}
\caption[]{Angular distribution of muons in CMS recorded with and without circulating beam. Also shown is the distribution for simulated beam-halo events.}
\label{fig:BeamHcosmic}
\end{figure}

The detectors and triggers that were used in September 2008 were sensitive to cosmic-ray muons as well as to beam-halo particles when a requirement of a signal from the beam pick-ups was not made. The presence of beam-halo and cosmic-ray signals in the data is illustrated in \Fref{fig:BeamHcosmic} which shows the angular distribution of muons reconstructed in CMS.  The shape of the cosmic-ray distribution, which has a broad peak centred around 0.3--0.4 radians, is known from data collected without beam. The peak at low angles matches well with the distribution for simulated beam-halo particles.

\begin{figure}[ht]
\centering\includegraphics[width=.8\linewidth]{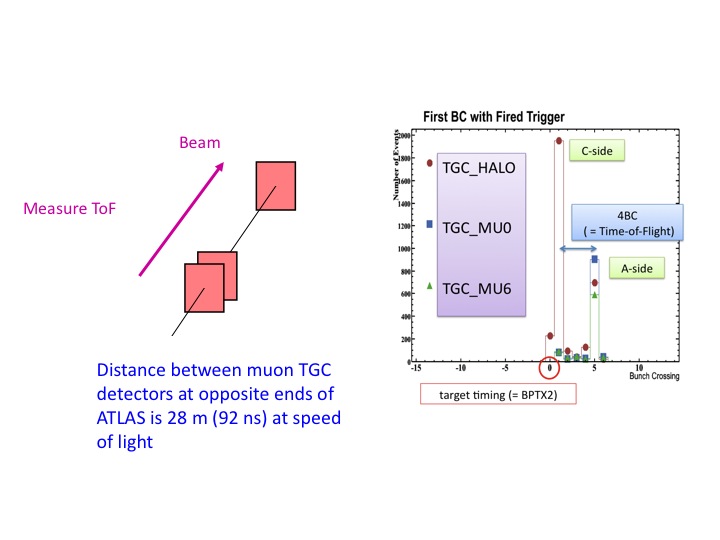}
\caption[]{Time of flight of beam-halo muons in ATLAS (one BC is 25 ns)}
\label{fig:TOF_TGC}
\end{figure}

Before concluding, the author would like to show another example of a study with single-beam data. Using a timing set-up in the end-cap muon trigger that would be appropriate for colliding-beam operations, in which the muons emerge from the centre of the apparatus, the distribution shown in the right-hand part of  \Fref{fig:TOF_TGC} was obtained. The two peaks separated by four bunch crossings, i.e., 4~$\times$~25~ns, correspond to triggers seen in the two ends of the detector system. This is consistent within the resolution with the time of flight of the beam-halo particles that may trigger the experiment on the upstream or downstream sides of the detector. As indicated in the left-hand part of the figure, this is reminiscent of the very simple example that was introduced early on in the lectures, see  \Figure~\ref{fig:f1}.

\section{Concluding remarks}

It is hoped that these lectures have succeeded in giving some insight
into the challenges of building T/DAQ systems for HEP
experiments. These include challenges connected with the physics
(inventing algorithms that are fast, efficient for the physics of
interest, and that give a large reduction in rate), and challenges in
electronics and computing. It is also hoped that the lectures have
demonstrated how the subject has evolved to meet the increasing
demands, \eg of LHC compared to LEP, by using new ideas based on new
technologies.

\section*{Acknowledgements}

The author would like to thank the local organizing committee for their wonderful
hospitality during his stay in Colombia. In particular, he would like
to thank Marta Losada and Enrico Nardi who, together, created such
a wonderful atmosphere between all the participants, staff and
students alike.

The author would like to thank the following people for their help and
advice in preparing the lectures and the present notes: Bob Blair,
Helfried Burckhart, Vincenzo Canale, Philippe Charpentier, Eric
Eisenhandler, Markus Elsing, Philippe Farthouat, John Harvey, Andreas Hoecker, Jim
Linnerman, Claudio Luci, Jordan Nash, Thilo Pauly, and Wesley Smith.

\end{document}